\documentclass[12pt]{article}
\usepackage{amsmath, amsthm, mathrsfs, amssymb, amsfonts, mathabx}
\usepackage{soul}
\usepackage{dsfont}
\usepackage{booktabs}
\usepackage[authoryear]{natbib}
\usepackage{times, latexsym}
\usepackage{color}
\usepackage{comment}
\usepackage{float}
\usepackage{setspace}
\usepackage{graphicx, epsfig}
\usepackage{subfigure}
\DeclareGraphicsExtensions{.eps,.pdf,.png,.jpg}
\usepackage{bbm}
\usepackage{mathtools}
\usepackage{framed}
\usepackage{fancybox}
\usepackage{rotating, threeparttable, booktabs, caption, dcolumn, pdflscape}
\usepackage{multirow}
\usepackage[titletoc, title]{appendix}
\newcolumntype{d}[1]{D..{#1}}

\usepackage[backref=page,
colorlinks=true,
citecolor=blue,
linkcolor=blue,
urlcolor=blue,
bookmarksnumbered=true,
bookmarksopen=true,
breaklinks=true,
pdfstartview=FitH
]{hyperref}
\renewcommand*\backref[1]{\ifx#1\relax \else [#1] \fi}

\usepackage{algorithm}
\usepackage{algpseudocode}

\algnewcommand\algorithmicassume{\textbf{Assume:}}
\algnewcommand\Assume{\item[\algorithmicassume]}
\algnewcommand{\LineComment}[1]{\State \(\triangleright\) #1}

\usepackage{listings}
		    \def\bE{\mathbb{E}}

\renewcommand{\hat}{\widehat}

\renewcommand{\bar}{\overline}

\def\prob {{\rm Pr}}

\allowdisplaybreaks[4]

\title{A Unified Framework for Density Estimation under Right-Censored Point-Centred Quarter Sampling}

\author{Wenzhe Huang$^{1}$, Guochun Shen$^{2}$, Dingliang Xing$^{2,3,}$\footnote{Corresponding author: email: dlxing@des.ecnu.edu.cn}, Jiangyan Zhao$^{4,}$\footnote{The authors' names are listed in alphabetical order to reflect equal contribution.}}

\date{}

\begin{document}
	
	\maketitle
	
	\vspace{-0.5cm}
	
	\noindent $^1$School of Mathematical Sciences, East China Normal University,  Shanghai 200241, China\\
	$^2$Zhejiang Tiantong Forest Ecosystem National Observation and Research Station, School of Ecological and Environmental Sciences, East China Normal University, Shanghai 200241, China\\
	$^3$Zhejiang Zhoushan Island Ecosystem Observation and Research Station, School of Ecological and Environmental Sciences, East China Normal University, Shanghai 200241, China\\
	$^4$KLATASDS-MOE, School of Statistics, East China Normal University,  Shanghai 200062, China
	
	\section*{DATA AVAILABILITY STATEMENT}
	
	No new data were collected for this study. This research utilized two publicly available datasets previously published by \citet{condit2019complete} and \citet{orwig2022harvard}. The custom R code used in our analysis is available on Zenodo: \url{https://doi.org/10.5281/zenodo.20390381}. The companion R package \texttt{TruncatedPCQM} is available on CRAN at \url{https://cran.r-project.org/package=TruncatedPCQM}.
	
	\newpage
	
	\begin{abstract}
		While the point-centred quarter method (PCQM) is widely used for density estimation, existing methods for handling right-censored data from truncated search radii rely primarily on a Poisson model assuming complete spatial randomness (CSR), leaving a critical gap for spatially aggregated populations.
		To address this limitation, we develop a unified likelihood- and moment-based framework for right-censored point-centred quarter sampling under both Poisson and negative binomial distribution (NBD) models. In particular, the proposed NBD-based estimators explicitly account for spatial aggregation and censoring simultaneously, extending distance-based inference beyond the CSR setting. Extensive simulations and applications to fully mapped forest plots reveal that the NBD-based MLE delivers the most robust overall performance across diverse ecological scenarios. Across more than 100 species from fully mapped forest plots, the proposed NBD-based MLE approximately reduced absolute relative bias by a median of 0.10 compared with existing censored estimators, representing a relative improvement of over 30\%. Ultimately, our framework provides a rigorously validated and practically useful toolkit for analysing censored point-to-tree distance data.
	\end{abstract}

	\noindent \textbf{Keywords:} complete spatial randomness; maximum likelihood estimation; point-centred quarter method; population density estimation; spatial aggregation; truncated sampling
	
	\section{Introduction}\label{sec:Intro}
	
	Population density is a fundamental ecological quantity that underpins inference on population dynamics, community structure, biodiversity patterns, and conservation planning \citep{cogbill2023surveyor,santini2024tetradensity}. 
	Reliable density estimates are therefore essential not only for ecological understanding, but also for monitoring and management. 
	For sessile organisms such as trees, distance-based sampling methods are especially attractive because they can greatly reduce field effort relative to quadrat-based surveys while remaining simple to implement and interpret. 
	Among these methods, the point-centred quarter method (PCQM) has been widely used in forest ecology and vegetation surveys for decades \citep{cottam1956use,mitchell2023quantitative,mienna2024quantifying}. 
	Recent syntheses and comparative studies further show that plotless distance-based estimators remain widely used in ecology, but their performance depends strongly on estimator choice, underlying spatial pattern, and survey design \citep{cogbill2018retrospective, notarangelo2023performance, pommerening2024distance}.
	
	The statistical basis of PCQM is well established under complete spatial randomness (CSR), typically represented by a homogeneous Poisson point process \citep{cottam1953some,morisita1954estimation}. 
	Under CSR, point-to-tree distances admit simple moment relationships that lead to closed-form estimators of population density. 
	In ecological applications, however, the CSR assumption is often unrealistic. 
	Spatial aggregation is common because of dispersal limitation, habitat heterogeneity, disturbance history, and biotic interactions. 
	When such clustering is present, Poisson-based PCQM estimators are known to systematically underestimate true density \citep{pollard1971distance,he1997distribution}. 
	This limitation has motivated a series of alternative estimators, including recent approaches based on the negative binomial distribution (NBD), which provide a more flexible parametric framework for accommodating aggregation and have shown substantially improved robustness for uncensored point-to-tree distance data \citep{shen2020distance,Stoklosa2022NBD}. 
	More broadly, recent work on distance-based sampling continues to highlight both the practical efficiency of these methods and their sensitivity to the spatial structure of the populations being sampled \citep{cogbill2018retrospective, pommerening2024distance}.
	
	A further complication, however, arises from the way PCQM is implemented in the field. 
	In many surveys, observers use a maximum search radius to maintain efficiency, improve safety, or accommodate restricted visibility in dense vegetation or complex terrain \citep{levine2017evaluating,cogbill2023surveyor}. 
	When no individual is encountered within this prespecified radius, the corresponding observation is right-censored. 
	Such censoring is not a minor technical detail: low density and spatial aggregation often coincide — the norm demonstrated by \citet{wiegand2025latitudinal}, who show that species with low abundance are more aggregated.
	Existing corrections were developed largely under CSR \citep{warde1981correction,dahdouh2006empirical}, and their performance can deteriorate markedly when applied to aggregated populations \citep{mitchell2023quantitative}. 
	Thus, although NBD-based estimators are now available for uncensored distance sampling \citep{shen2020distance}, there is still no unified framework for analysing right-censored PCQM data under both random and aggregated spatial structure.
	
	Here, we address this gap by developing a unified framework for density estimation under right-censored point-centred quarter sampling. 
	We focus on two core spatial models that anchor the existing literature: the Poisson model for populations under CSR and the NBD model for aggregated populations. 
	Under the Poisson framework, we extend classical moment-based estimators to account for right-censoring and derive the corresponding maximum likelihood estimator. 
	Under the NBD framework, we develop new moment-based and likelihood-based estimators that jointly accommodate aggregation and censoring. 
	Together, these methods provide, to our knowledge, the first systematic treatment of censored PCQM data that is valid under both random and aggregated spatial structures.
	By doing so, this study links a common field constraint with ecologically realistic spatial models and establishes a practical inferential basis for analysing censored point-to-tree distance data.
	
	We assess the proposed estimators through extensive simulations spanning a wide range of censoring levels, aggregation strengths, and sampling settings, and through applications to fully mapped forest plot data. 
	
	\section{Existing Estimators}\label{sec:background}
	
	We first review population density estimators for complete (i.e., uncensored) point-to-individual distance data, and then summarize existing corrections for right‑censored sampling. 
	These form the methodological basis for our subsequent extension to right‑censored point‑centred quarter sampling. 
	Consider a standard point-centred quarter sampling design in which $n$ sampling locations are independently selected within a study region. 
	At each location, the surrounding space  is partitioned into $q$ equal-angle sectors, and the distance $r_{\ell ij}$ from sampling point $i$ ($i=1,\dots,n$) to the $\ell$th nearest individual ($\ell=1,2,\dots$) in sector $j$ ($j=1,\dots,q$) is recorded. 
	This yields a total of $nq$ observed distances, denoted by $\{r_m:m=1,\dots,nq\}$ for notational simplicity.
	
	The form and performance of distance-based density estimators depend directly on assumptions about the underlying spatial point process. 
	In what follows, we summarize the principal estimators under two canonical models commonly used in ecological applications: complete spatial randomness, represented by the Poisson model, and spatial aggregation, represented by the negative binomial distribution model.
	
	\subsection{Complete Data: Poisson (CSR) Case}
	
	Under complete spatial randomness (CSR), individuals are assumed to follow a homogeneous Poisson point process. 
	The Poisson model can also be regarded as a limiting case of the NBD model as the aggregation parameter $k \to \infty$ \citep{eberhardt1967developments}. 
	Under CSR, the probability density function of the distance $R$ to the $\ell$th nearest individual within each of the $q$ equal-angle sectors is
	\begin{equation} \label{eq:CSR_complete_pdf}
		f(r;\lambda) = \frac{2\pi \lambda r}{q} \frac{(\lambda \pi r^2/q)^{\ell-1}}{(\ell-1)!} \exp\!\left(-\lambda \pi r^2/q\right),
	\end{equation}
	where $\lambda$ denotes population density.
	
	Several classical density estimators follow from this distribution. Following the classification of \citet{cogbill2018retrospective}, two widely used forms are the Cottam-type and Pollard-type estimators. 
	The Cottam-type estimator, originally generalized by \citet{morisita1954estimation}, is
	\begin{equation}\label{eq:Cottam}
		\hat{\lambda}_{C} = \frac{q \ell}{4 \left[ \frac{1}{nq} \sum_{m=1}^{nq} r_m \right]^2},
	\end{equation}
	which depends on the first moment of the observed distances. 
	The Pollard-type estimator \citep{morisita1957new} instead uses the second moment and is given by
	\begin{equation} \label{eq:Pollard}
		\hat{\lambda}_{P} = \frac{q (n q \ell - 1)}{\pi \sum_{m=1}^{nq} r_m^{2}}.
	\end{equation}
	
	A likelihood-based estimator can also be obtained directly from \eqref{eq:CSR_complete_pdf}. The resulting maximum likelihood estimator (MLE) of $\lambda$ is
	\begin{equation} \label{eq:MLE_CSR}
		\hat{\lambda}_{\text{MLE}} = \frac{n q^2 \ell}{\pi \sum_{m=1}^{nq} r_m^{2}}.
	\end{equation}
	Although this estimator arises naturally from the likelihood, it is typically less attractive in practice than the Pollard-type estimator because of its bias properties \citep{pollard1971distance}.
	
	\subsection{Complete Data: Negative Binomial (Aggregation) Case}
	
	The usefulness of Poisson-based estimators depends on the CSR assumption, which is often violated in ecological systems where individuals are spatially aggregated. 
	Such aggregation may arise from dispersal limitation, habitat heterogeneity, or biotic interactions, and in these settings CSR-based estimators can substantially underestimate true density \citep{bryant2005forest}. 
	
	Early attempts to relax the CSR assumption include the estimators proposed by \citet{morisita1957new}, which assume local randomness within sectors even when the broader spatial pattern is aggregated. Morisita's two estimators based on point-to-tree distances are
	\begin{equation}\label{eq:m1}
		\hat{\lambda}_{m1} = \frac{\ell - 1}{\pi n} \sum_{m=1}^{nq} \frac{1}{r_{m}^{2}},
	\end{equation}
	which requires $\ell > 1$, and
	\begin{equation}\label{eq:m2}
		\hat{\lambda}_{m2} = \frac{\ell q - 1}{\pi n} \sum_{i=1}^{n} \frac{q}{\sum_{j=1}^{q} r_{ij}^{2}}.
	\end{equation}
	Comprehensive evaluations by \citet{cogbill2018retrospective} show that these estimators remain among the best-performing classical plotless methods when CSR is violated.
	
	A more explicit treatment of aggregation was introduced by \citet{shen2020distance}, who modeled point-to-tree distances under the negative binomial distribution. 
	The NBD has long been recognized as a flexible model for aggregated spatial patterns \citep{ZILLIO2010modelling,Condit2018community}, but earlier uses in distance sampling often relied on numerical inversion or simulation-based procedures \citep{eberhardt1967developments,gao2013detecting}. 
	In contrast, \citet{shen2020distance} derived closed-form moment-based estimators, making density estimation under aggregation both direct and computationally efficient.
	
	Under the NBD model, the probability density function of the distance $R$ to the $\ell$th nearest individual in each sector is
	\begin{equation}\label{eq:NBD_complete_pdf}
		g(r;\lambda,k) =\frac{2 (\pi \lambda q^{-1})^{\ell} r^{2\ell - 1} \Gamma(\ell + k)}{k^\ell \Gamma(k) \Gamma(\ell)}\left(1 + \frac{\pi \lambda q^{-1} r^{2}}{k} \right)^{-\ell - k},
	\end{equation}
	where $\lambda$ is population density and $k$ controls the strength of spatial aggregation \citep{shen2020distance}. The $u$th moment of $R$ is
	\[
	\bE[R^u] =\left( \frac{kq}{\pi \lambda} \right)^{u/2}\frac{\Gamma\left(\ell + \frac{u}{2}\right)\Gamma\left(k - \frac{u}{2}\right)}{\Gamma(\ell)\Gamma(k)},\qquad -2\ell < u < 2k.
	\]
	When $u=-2$, this expression reduces to Morisita's first estimator \eqref{eq:m1}, as shown by \citet{eberhardt1967developments}. 
	When $u=2$, the second moment becomes $\bE[R^2] = \frac{k \ell q}{\pi \lambda (k - 1)}$, which requires $k > 1$. 
	This restriction limits the use of second-moment-based estimators in strongly aggregated populations, although such estimators remain convenient and effective for moderately clustered communities.
	
	By combining moment conditions based on $\bE[R^{-1}]$, $\bE[R]$, and $\bE[R^2]$, \citet{shen2020distance} obtained the NBD-based density estimator
	\begin{equation}\label{eq:shen}
		\hat{\lambda}_{n} =\frac{q(2\ell - 1) \sum_{m=1}^{nq} r_{m}^{-1}}{\pi \sum_{m=1}^{nq} r_{m}}-\frac{nq^2 \ell}{\pi \sum_{m=1}^{nq} r_{m}^2},
	\end{equation}
	together with the corresponding estimator of the aggregation parameter
	\begin{equation}\label{eq:shen_k}
		\hat{k}_{n} =1 -\frac{\left(\sum_{m=1}^{nq} r_m\right)\ell}{\left(\sum_{m=1}^{nq} r_m^{-1}\right)\left(\sum_{m=1}^{nq} r_m^2\right)\frac{1-2\ell}{nq}+ \left(\sum_{m=1}^{nq} r_m\right)\ell}.
	\end{equation}
	
	Likelihood-based inference under the NBD model is obtained by maximizing
	\begin{equation}\label{eq:completenbd}
		\mathcal{L}(\lambda, k) = \prod_{m=1}^{nq} g(r_m;\lambda,k).
	\end{equation}
	with respect to $\lambda$ and $k$. 
	Because closed-form solutions are unavailable, numerical optimization is required. These likelihood-based estimators provide an important benchmark for the complete-data setting and motivate our extension to right-censored distance data in the next section.
	
	\subsection{Right-Censored Data and Existing Corrections}\label{sec:Rightcensored}
	
	In practical field surveys, distance sampling is often constrained by limited time, manpower, or visibility, making a fixed maximum search radius $C$ necessary. 
	When distances to the $\ell$th nearest individual are recorded within each of $q$ sectors, any sector in which fewer than $\ell$ individuals are detected within radius $C$ yields a right-censored observation, that is, $R > C$. 
	Such censored sectors are common in low-density stands or heterogeneous habitats, and ignoring them can induce substantial positive bias in density estimation \citep{warde1981correction}.
	
	Let $n_0$ denote the number of censored sectors, let $\{r'_t:t=1,2,\dots,nq-n_0\}$ denote the fully observed distances, and let $p_0=n_0/(nq)$ be the proportion of censored observations. 
	Throughout, we consider the standard PCQM design \citep{cogbill2018retrospective}, in which only the distance to the $\ell$th nearest neighbour is recorded in each of the $nq$ sectors. 
	Existing corrections for censored PCQM data have been developed primarily under CSR \citep{mitchell2023quantitative}. 
	For the special case $\ell=1$, \citet{warde1981correction} proposed a correction based on the censored proportion $p_0$, yielding
	\begin{equation}\label{eq:WP}
		\hat{\lambda}_{\text{WP}} =\frac{q}{\pi \left( \frac{1}{nq-n_0} \sum_{t=1}^{nq-n_0} r'_t \right)^2}\cdot\frac{\left[\gamma\left( \frac{3}{2}, -\ln(p_0) \right)\right]^2}{(1-p_0)^2},
	\end{equation}
	where $\gamma(a,x)=\int_0^x t^{a-1}e^{-t}\,dt$ is the lower incomplete gamma function. 
	Subsequently, \citet{dahdouh2006empirical} proposed a simpler empirical correction,
	\begin{equation}\label{eq:DK}
		\hat{\lambda}_{\text{DK}} =\frac{q}{4 \left( \frac{1}{nq-n_0} \sum_{t=1}^{nq-n_0} r'_t \right)^2}(1-p_0),
	\end{equation}
	which performs reasonably well in practice despite lacking formal theoretical justification.

	\section{Improved Estimators for Right-censored Data}\label{sec:improved}
	
	We now extend the complete-data estimators reviewed in Sections~\ref{sec:background} to the right-censored setting introduced in Section~\ref{sec:Rightcensored}. As in the previous section, we develop the results under two canonical spatial models: the Poisson model for populations under CSR and the NBD model for aggregated populations.
	
	\subsection{Poisson (CSR) Case}
	
	Using the notation and existing corrections introduced in Section~\ref{sec:background}, we now develop improved estimators under CSR that explicitly incorporate right-censoring. 
	
	To generalize moment-based density estimation under CSR, we replace the usual empirical moments by censoring-adjusted moments. Specifically, for any real number $u$, we replace the complete-data moment $\frac{1}{nq}\sum_{m=1}^{nq} r_m^{u}$ by
	\[
	M_u =\frac{\frac{1}{nq} \sum_{t=1}^{nq-n_0} r_t^{'u}\cdot\Gamma\!\left(\ell + \frac{u}{2}\right)}{\gamma\!\left(\ell + \frac{u}{2}, \hat{m}_C \right)},
	\]
	where $\hat{m}_C$ solves $\gamma(\ell, \hat{m}_C)/\Gamma(\ell) = 1 - p_0$. 
	When the maximum search radius $C$ is sufficiently large, $M_u$ reduces to the standard empirical moment; see Supplement Section \S 1.
	
	Replacing the complete-data moments in \eqref{eq:Cottam} and \eqref{eq:Pollard} with their censoring-adjusted counterparts yields the censored Cottam-type estimator
	\begin{equation}\label{eq:Cottam_censored}
		\hat{\lambda}_{C}^{(c)} = \frac{q \ell}{4(M_1)^2},
	\end{equation}
	and the censored Pollard-type estimator
	\begin{equation}\label{eq:Pollard_censored}
		\hat{\lambda}_{P}^{(c)} = \frac{n q \ell - 1}{\pi n M_2}.
	\end{equation}
	When $\ell=1$, \eqref{eq:Cottam_censored} reduces exactly to the Warde--Petranka estimator in \eqref{eq:WP}. 
	These estimators therefore extend classical CSR-based moment methods while explicitly accounting for right-censoring.
	
	Likelihood-based estimation under CSR can be extended in a similarly direct way. 
	For $nq-n_0$ observed distances $\{r'_t\}$ and $n_0$ censored sectors, the likelihood is
	\begin{equation}\label{eq:likelihood_CSR_censored}
		\mathcal{L}(\lambda) =\prod_{t=1}^{nq-n_0} f(r'_t;\lambda)\left[1 - F(C;\lambda)\right]^{n_0},
	\end{equation}
	where $f(\cdot)$ is given in \eqref{eq:CSR_complete_pdf} and
	\[
	F(C;\lambda) =\frac{\gamma\left(\ell, \pi \lambda C^2 / q\right)}{\Gamma(\ell)}.
	\]
	The maximum likelihood estimator $\hat{\lambda}_{\text{MLE}}^{(c)}$ is obtained numerically for general $\ell$. 
	In the special case $\ell=1$, it has the closed-form expression
	\begin{equation}\label{eq:MLE_censored_ell1}
		\hat{\lambda}_{\text{MLE}}^{(c)}=
		\frac{q(nq-n_0)}
		{\pi\left(\sum_{t=1}^{nq-n_0} r_t'^2+n_0C^2\right)},
	\end{equation}
	which reduces to \eqref{eq:MLE_CSR} when $n_0=0$.
	
	\subsection{Negative Binomial (Aggregation) Case}
	
	We next extend the NBD framework to right-censored distance sampling, thereby allowing spatial aggregation and truncation to be handled simultaneously. 
	The main idea is to construct adjusted sample moments in which censored distances are replaced by their conditional expectation given the truncation threshold. 
	For computational tractability and robust performance across aggregation levels, we approximate this conditional expectation under CSR limit $k=\infty$. 
	Supplement Section \S 2 shows that the resulting approximation error is negligible under moderate censoring and vanishes as $k$ increases.
	
	Accordingly, we define the adjusted moment
	\[
	\widehat{\bE[R^u]} =\frac{1}{nq}\left[\sum_{t=1}^{nq-n_0} r_t^{'u}+n_0 \,\mu_u^{(c)}\right],
	\]
	where
	\[
	\mu_u^{(c)} = \bE\!\left(R^u \mid R>C;\,k=\infty,\lambda=\lambda_{\text{init}}\right) =\left( \frac{q}{\pi \lambda_{\text{init}}} \right)^{u/2}\frac{\Gamma\!\left(\ell + \frac{u}{2}, \frac{\pi \lambda_{\text{init}} C^2}{q}\right)}{\Gamma\!\left(\ell, \frac{\pi \lambda_{\text{init}} C^2}{q}\right)}.
	\]
	Here, $\Gamma(a,x)$ denotes the upper incomplete gamma function, and $\lambda_{\text{init}}$ is an initial density estimate obtained from any censored-data estimator, for example \eqref{eq:DK}, \eqref{eq:Cottam_censored}, \eqref{eq:Pollard_censored}, or the Poisson-based censored MLE.
	
	Using these adjusted moments, the NBD-based estimator in \eqref{eq:shen} extends to
	\begin{equation}\label{eq:shen_censored}
		\hat{\lambda}_{n}^{(c)} =\frac{q(2\ell - 1)\widehat{\bE[R^{-1}]}}{\pi \widehat{\bE[R]}}-\frac{q \ell}{\pi \widehat{\bE[R^2]}},
	\end{equation}
	and Morisita's first estimator in \eqref{eq:m1} becomes
	\begin{equation}\label{eq:m1_censored}
		\hat{\lambda}_{m1}^{(c)} =\frac{q(\ell - 1)}{\pi} \widehat{\bE[R^{-2}]}.
	\end{equation}
	These extensions make it possible to apply NBD-based methods, which are known to perform well under aggregation, to censored distance data commonly encountered in field surveys.
	
	Finally, likelihood-based inference under the NBD model can also be extended to right-censored data. 
	For $nq-n_0$ observed distances and $n_0$ censored sectors, the likelihood is
	\begin{equation}\label{eq:likelihood_NBD_censored}
		\mathcal{L}(\lambda,k) =\prod_{t=1}^{nq-n_0} g(r'_t;\lambda,k)\left[1 - F(C;\lambda,k)\right]^{n_0},
	\end{equation}
	where $g(\cdot)$ is given in \eqref{eq:NBD_complete_pdf}. 
	The cumulative distribution function is
	\[
	F(C;\lambda,k) = I_w(\ell,k),\qquad w = \frac{\pi \lambda C^2}{\pi \lambda C^2 + qk},
	\]
	where $I_w(\cdot,\cdot)$ denotes the regularized incomplete beta function. The maximum likelihood estimators $\hat{\lambda}_{\text{n,MLE}}^{(c)}$ and $\hat{k}_{\text{n,MLE}}^{(c)}$ are obtained numerically by maximizing the corresponding log-likelihood.
	
	A comprehensive list of all estimators considered in this paper is provided in Table~\ref{tab:estimators_summary}.
	
	\begin{table}[!t]
		\centering
		\caption{Summary of density and parameter estimators considered in this study.}
		\label{tab:estimators_summary}
		\small
		\begin{tabular}{@{}l >{\raggedright\arraybackslash}p{10.0cm} >{\raggedright\arraybackslash}p{3.5cm}@{}}
			\toprule
			Notation & Definition & Source \\
			\midrule
			\multicolumn{3}{c}{\textit{Complete data}} \\
			$\hat{\lambda}_C$ 
			& Cottam-type estimator, Eq.~\eqref{eq:Cottam} 
			& \citep{morisita1954estimation} \\
			$\hat{\lambda}_P$ 
			& Pollard-type estimator, Eq.~\eqref{eq:Pollard} 
			& \citep{morisita1957new} \\
			$\hat{\lambda}_{\text{MLE}}$ 
			& Poisson maximum likelihood estimator, Eq.~\eqref{eq:MLE_CSR} 
			& \citep{pollard1971distance} \\
			$\hat{\lambda}_{m1}$ 
			& Morisita's first estimator, Eq.~\eqref{eq:m1} 
			& \citep{morisita1957new} \\
			$\hat{\lambda}_{m2}$ 
			& Morisita's second estimator, Eq.~\eqref{eq:m2} 
			& \citep{morisita1957new} \\
			$\hat{\lambda}_n$, $\hat{k}_n$ 
			& NBD-based moment estimator, Eq.~\eqref{eq:shen}, Eq.~\eqref{eq:shen_k}  
			& \citep{shen2020distance} \\
			$\hat{\lambda}_{\text{n,MLE}}$, $\hat{k}_{\text{n,MLE}}$ 
			& NBD maximum likelihood estimators (complete data), Eq.~\eqref{eq:completenbd} 
			& \citep{gao2013detecting} \\
			\midrule
			\multicolumn{3}{c}{\textit{Censored data}} \\
			$\hat{\lambda}_{\text{WP}}$ 
			& Warde--Petranka correction (special case of Eq.~\eqref{eq:Cottam_censored}), Eq.~\eqref{eq:WP} 
			& \citep{warde1981correction} \\
			$\hat{\lambda}_{\text{DK}}$ 
			& Dahdouh-Guebas--Koedam correction, Eq.~\eqref{eq:DK} 
			& \citep{dahdouh2006empirical} \\
			$\hat{\lambda}_{C}^{(c)}$ 
			& Censored Cottam-type estimator, Eq.~\eqref{eq:Cottam_censored} 
			& This study \\
			$\hat{\lambda}_{P}^{(c)}$ 
			& Censored Pollard-type estimator, Eq.~\eqref{eq:Pollard_censored} 
			& This study \\
			$\hat{\lambda}_{\text{MLE}}^{(c)}$ 
			& Censored Poisson MLE, Eq.~\eqref{eq:likelihood_CSR_censored} 
			& This study \\
			$\hat{\lambda}_{m1}^{(c)}$ 
			& Censored Morisita's first estimator, Eq.~\eqref{eq:m1_censored} 
			& This study \\
			$\hat{\lambda}_{n}^{(c)}$ 
			& Censored NBD-based moment estimator, Eq.~\eqref{eq:shen_censored} 
			& This study \\
			$\hat{\lambda}_{\text{n,MLE}}^{(c)}$, $\hat{k}_{\text{n,MLE}}^{(c)}$ 
			& Censored NBD maximum likelihood estimators, Eq.~\eqref{eq:likelihood_NBD_censored} 
			& This study \\
			\bottomrule
		\end{tabular}
	\end{table}

	\section{Study Design and Evaluation Criteria}
	
	We evaluated the proposed estimators for right-censored point-centred quarter sampling using both simulated spatial point patterns and fully mapped forest plots. 
	The simulation study was designed to span a broad range of spatial structures commonly encountered in ecological populations, from complete spatial randomness to strong aggregation, whereas the empirical analyses were intended to assess estimator behaviour under realistic forest conditions.
	
	\subsection{Simulation Study}
	
	We considered two broad classes of spatial point processes.
	
	\paragraph{Complete spatial randomness (CSR).}
	
	CSR populations were generated from homogeneous Poisson point processes, representing the classical assumption underlying traditional plotless distance estimators \citep{Thomas1949}. 
	These simulations provide a baseline for evaluating censored estimators when the Poisson model is correctly specified. 
	We considered three population densities, $\lambda_{\text{CSR}} = 0.005$, $0.01$, and $0.05$ individuals per unit area, spanning sparse to moderately dense populations representative of ecological field surveys.
	
	Under CSR, we compared four Poisson-based estimators:
	\begin{enumerate}
		\item the Dahdouh-Guebas and Koedam estimator, $\hat{\lambda}_{\text{DK}}$ \eqref{eq:DK}, applicable when only the nearest neighbor ($\ell = 1$) is recorded;
		\item the censored Cottam-type estimator, $\hat{\lambda}_{C}^{(c)}$ \eqref{eq:Cottam_censored};
		\item the censored Pollard-type estimator, $\hat{\lambda}_{P}^{(c)}$ \eqref{eq:Pollard_censored};
		\item the maximum likelihood estimator under CSR, $\hat{\lambda}_{\text{MLE}}^{(c)}$, based on the censored likelihood in \eqref{eq:likelihood_CSR_censored}.
	\end{enumerate}
	This comparison isolates the relative efficiency and robustness of Poisson-based estimators when their underlying assumptions are satisfied.
	
	\paragraph{Aggregated (non-CSR) populations.}
	
	To represent spatial aggregation arising from ecological processes such as limited dispersal, clonal growth, and habitat heterogeneity, we simulated Thomas cluster processes \citep{Thomas1949}. 
	The overall population density was fixed at $\lambda_{\text{agg}} = 0.05$, matching the highest CSR density, while aggregation strength was controlled through the cluster scale parameter $\sigma$. 
	Parent intensity was set to $\kappa = \lambda_{\text{agg}} / \mu$ with $\mu = 5$ offspring per parent, and $\sigma$ varied from 1.0 to 5.5 units in increments of 0.5. 
	Smaller values of $\sigma$ generate strongly aggregated patterns, whereas larger values approach spatial randomness.
	
	For aggregated populations, we evaluated seven estimators spanning both modeling frameworks: the four Poisson-based estimators \eqref{eq:DK}, \eqref{eq:Cottam_censored}, \eqref{eq:Pollard_censored}, and \eqref{eq:likelihood_CSR_censored}, together with three estimators derived under the NBD framework:
	\begin{enumerate}
		\item the censored Shen-type estimator, $\hat{\lambda}_{n}^{(c)}$ \eqref{eq:shen_censored};
		\item the censored Morisita-type estimator, $\hat{\lambda}_{m1}^{(c)}$ \eqref{eq:m1_censored};
		\item the NBD-based maximum likelihood estimator, $\hat{\lambda}_{\text{n,MLE}}^{(c)}$ \eqref{eq:likelihood_NBD_censored}.
	\end{enumerate}
	This comparison allows us to assess the robustness of NBD-based methods when spatial aggregation is present and to compare their performance with CSR-based estimators under model misspecification.

	For the case $\ell = 1$, we include the  estimators $\hat{\lambda}_{\text{DK}}$ \eqref{eq:DK} and estimator $\hat{\lambda}_{C}^{(c)}$ \eqref{eq:Cottam_censored} as benchmarks in the main-text figures alongside the three NBD-based estimators (Figure~\ref{fig:nbd_estimators}). 
	This allows direct comparison of our proposed methods with the established corrections of \citet{warde1981correction} and \citet{dahdouh2006empirical}. 
	Complete results for all four Poisson-based estimators across all values of $\ell$ are provided in Supplement Section \S3.
	
	For each combination of process parameters, we generated 10 independent point-pattern realizations in a $600 \times 600$ unit square study window using the \texttt{spatstat} package in \textsf{R} \citep{baddeley2015spatstat}. Within each realization, we implemented a point-centred quarter sampling design with $n = 120$ sampling points, $q = 4$ quadrants per point, and nearest-neighbour orders $\ell = 1, 2,$ and $3$. To reflect realistic field constraints, we imposed a maximum search radius of $C = 10$ units and treated all distances exceeding this threshold as right-censored observations. Sampling locations were generated using Latin hypercube sampling via the \texttt{lhs} package \citep{carnell2026lhs} within a buffered study window, with a minimum distance of $C + 0.1$ units from the boundary to avoid edge effects \citep{OLSSON2014Latin}. For each point pattern, 200 independent sets of sampling locations were generated to quantify variability arising from sampling placement.
	
	\subsection{Empirical Forest Data}
	
	To assess estimator performance under realistic ecological conditions, we applied all seven estimators to two fully censused forest dynamics plots: the 50-ha Barro Colorado Island (BCI) plot in Panama \citep{condit2019complete} and the 35-ha Harvard Forest (HF) plot in Massachusetts, USA \citep{orwig2022harvard}. These datasets span a wide range of species abundances and spatial structures, including strong aggregation and habitat-driven heterogeneity.
	
	For each species with at least 500 individuals, we simulated PCQM sampling with $n = 120$ sampling points and $q = 4$ quadrants. This abundance threshold yielded 112 species for the BCI plot and 20 species for the Harvard Forest plot. We considered four maximum search radii ($C = 10$, 20, 30 and 40 m) and three nearest-neighbour orders ($\ell = 1, 2, 3$). Sampling points were generated using Latin hypercube sampling with a buffer of $C + 0.1$ m to avoid edge effects. For each species and parameter combination, 200 independent sampling designs were generated. True species densities were calculated directly from the full census, allowing the same performance metrics used in the simulation study to be evaluated under empirical forest conditions.
	
	\subsection{Performance Metrics}
	
	Estimator performance was evaluated using relative error metrics calculated across the 200 sampling replicates for each scenario. Let $\hat{\lambda}_s$ denote the estimated density from the $s$th replicate and let $\lambda_{\text{true}}$ denote the true population density, computed as the total number of individuals divided by the study area. Relative bias (rBias) measures systematic deviation,
	\[
	\text{rBias} = \frac{1}{\lambda_{\text{true}}} \left( \frac{1}{200} \sum_{s=1}^{200} \hat{\lambda}_s - \lambda_{\text{true}} \right),
	\]
	while relative root mean squared error (rRMSE) summarizes overall accuracy,
	\[
	\text{rRMSE} = \frac{1}{\lambda_{\text{true}}} \sqrt{ \frac{1}{200} \sum_{s=1}^{200} \left( \hat{\lambda}_s - \lambda_{\text{true}} \right)^2 }.
	\]
	Precision was quantified using the relative standard deviation (rSD),
	\[
	\text{rSD} = \frac{1}{\lambda_{\text{true}}} \sqrt{ \frac{1}{199} \sum_{s=1}^{200} \left( \hat{\lambda}_s - \bar{\hat{\lambda}} \right)^2 },
	\]
	where $\bar{\hat{\lambda}}$ is the mean estimated density across replicates.
	
	All simulations and analyses were conducted in \textsf{R}. 
	The companion \textsf{R} package \texttt{TruncatedPCQM} is available on CRAN at \url{https://cran.r-project.org/package=TruncatedPCQM}. 

	\section{Results}
	
	\subsection{Performance of the Estimators for Simulated Populations}
	
	\subsubsection{CSR Populations}
	
	For populations generated under the Poisson model, the estimators $\hat{\lambda}_{C}^{(c)}$, $\hat{\lambda}_{P}^{(c)}$, and $\hat{\lambda}_{\text{MLE}}^{(c)}$ all showed low relative bias, indicating similarly robust performance across population densities and nearest-neighbour orders. 
	By contrast, $\hat{\lambda}_{\text{DK}}$, which is applicable only when $\ell=1$, exhibited substantial positive bias at low densities, where censoring was most severe. 
	More generally, the relative performance of all estimators was strongly influenced by the resulting censored rate, defined as $n_0/(nq)$.

	\begin{table}[!t]
		\renewcommand{\arraystretch}{1.2}
		\setlength{\tabcolsep}{10pt}
		\centering
		\caption{Relative bias and precision of Poisson-based estimators for simulated complete spatial randomness populations.
			Values are presented as rBias (rSD). \textbf{Bolded} values indicate the smallest absolute relative bias for each scenario.
			The symbol ``---'' indicates that the estimator is not applicable for the given $\ell$.}
		\label{tab:csr_sim_results}
		\begin{tabular}{ccccccc}
			\toprule
			Intensity & $\ell$ & $\hat{\lambda}_{\text{MLE}}^{(c)}$ & $\hat{\lambda}_{C}^{(c)}$ & $\hat{\lambda}_{\text{DK}}$ & $\hat{\lambda}_{P}^{(c)}$ & Censored Rate \\
			\midrule
			& 1 & -0.0034 & \textbf{-0.0017} & 0.5812  & -0.0041 & 0.676 \\
			&  & (0.0796) &(0.0936) & (0.1385) & (0.0884) &  \\
			0.005 & 2 & -0.0065  & -0.1120  & --- & \textbf{-0.0018}  & 0.941 \\
			&  &  (0.1059) &  (0.1161) &  &  (0.1243) &  \\
			& 3 & -0.0220  & -0.1443   & --- & \textbf{0.0060}   & 0.993 \\
			&  & (0.1853) & (0.2253) &   &  (0.2550) &   \\
			\midrule
			& 1 & -0.0046 & \textbf{-0.0020}  & 0.4415  & -0.0043   & 0.458 \\
			&   &  (0.0619) &  (0.0737) &  (0.0958) &  (0.0689) &  \\
			0.010 & 2 & \textbf{0.0026}  & -0.1051  & --- & 0.0051   & 0.813 \\
			&   &  (0.0630) &  (0.0682) &   &  (0.0728) &  \\
			& 3 & \textbf{0.0004}  & -0.1430 & --- & 0.0039 & 0.954 \\
			&  & (0.0886) &  (0.0930) &   &  (0.1047) &   \\
			\midrule
			& 1 & \textbf{0.0005}  & 0.0011 & 0.0425   & -0.0009 & 0.020 \\
			&  & (0.0451) &  (0.0479) & (0.0476) &  (0.0455) &  \\
			0.050 & 2 & \textbf{-0.0012}   & -0.1119  & --- & -0.0017 & 0.098 \\
			&   &  (0.0328) &  (0.0309) &  & (0.0335) &  \\
			& 3 & \textbf{-0.0020}  & -0.1485   & --- & -0.0027  & 0.250 \\
			&  &  (0.0282) &  (0.0258) &  &  (0.0294) &   \\
			\bottomrule
		\end{tabular}
	\end{table}

	Table~\ref{tab:csr_sim_results} summarizes relative bias and precision, measured by the relative standard deviation, for the four Poisson-based estimators across the CSR simulation scenarios.

	\subsubsection{Non-CSR Populations}
	
\begin{figure}[!t]
		\centering
		\includegraphics[width=1\linewidth]{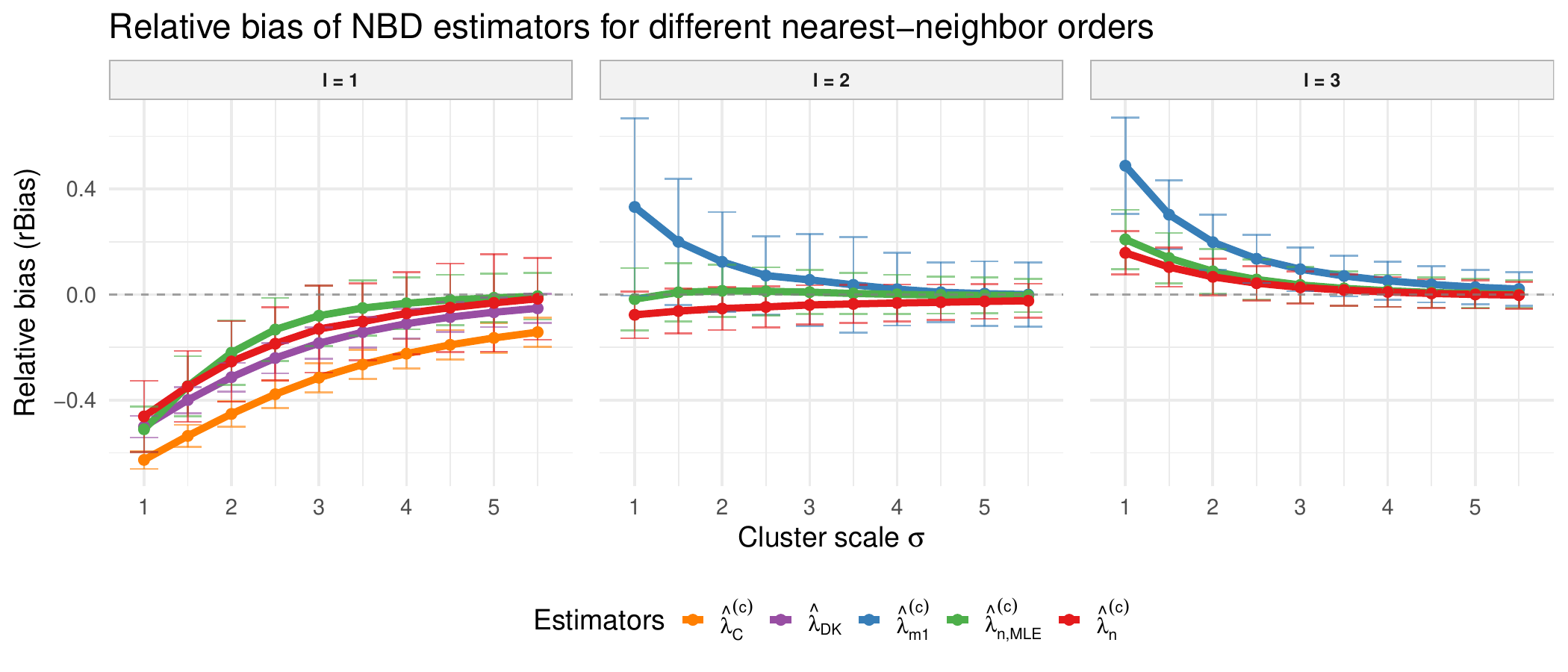}
		\caption{Relative bias (points and solid lines) of censored density estimators across a gradient of cluster scales ($\sigma$) for nearest-neighbour orders $\ell = 1, 2,$ and $3$. 
			In the $\ell=1$ panel, four estimators are shown: the NBD-based $\hat{\lambda}_{\text{n,MLE}}^{(c)}$ (green), $\hat{\lambda}_{n}^{(c)}$ (red), together with the Poisson-based $\hat{\lambda}_{\text{DK}}$ (purple) and $\hat{\lambda}_{C}^{(c)}$ (orange). 
			The estimator $\hat{\lambda}_{m1}^{(c)}$ (blue) is not applicable for $\ell=1$ and is therefore omitted.
			For $\ell=2$ and $\ell=3$, the three NBD estimators are displayed: $\hat{\lambda}_{\text{n,MLE}}^{(c)}$ (green), $\hat{\lambda}_{n}^{(c)}$ (red), and $\hat{\lambda}_{m1}^{(c)}$ (blue).
			Error bars represent $\pm$ one relative standard deviation. Simulations are based on a Thomas cluster process with intensity $\lambda = 0.05$ and censoring radius $C = 10$ m.}
		\label{fig:nbd_estimators}
	\end{figure}

	For aggregated populations generated from the Thomas process, the performance of the NBD-based censored estimators depended on both the aggregation scale ($\sigma$) and the nearest-neighbour order ($\ell$).
	Figure~\ref{fig:nbd_estimators} shows the relative bias of $\hat{\lambda}_{n}^{(c)}$, $\hat{\lambda}_{m1}^{(c)}$, and $\hat{\lambda}_{\text{n,MLE}}^{(c)}$ across a gradient of $\sigma$ values for $\ell=1,2,$ and $3$.

	When $\ell=1$ or $2$, $\hat{\lambda}_{\text{n,MLE}}^{(c)}$ was generally less biased than $\hat{\lambda}_{n}^{(c)}$ across most aggregation scales. 
	At $\ell=3$, the two estimators showed broadly comparable bias. 
	In terms of precision, $\hat{\lambda}_{\text{n,MLE}}^{(c)}$ and $\hat{\lambda}_{n}^{(c)}$ performed similarly across scenarios. 
	Both estimators consistently outperformed $\hat{\lambda}_{m1}^{(c)}$, which showed larger bias and substantially greater uncertainty.
	
	By contrast, the four Poisson-based estimators, all of which assume complete spatial randomness, exhibited negative bias across all aggregated scenarios, with the magnitude of bias increasing as aggregation became stronger. 
	Overall, these estimators were clearly outperformed by the NBD-based maximum likelihood estimator $\hat{\lambda}_{\text{n,MLE}}^{(c)}$ and the censored Shen-type estimator $\hat{\lambda}_{n}^{(c)}$ in relative bias (especially $\hat{\lambda}_{\text{DK}}$ and $\hat{\lambda}_{C}^{(c)}$ for $\ell = 1$).
	Full quantitative results for the four Poisson-based estimators are reported in Supplement Section \S3.

	\subsection{Performance on Empirical Forest Plot Data}
	
\begin{figure}[!t]
		\centering
		\includegraphics[width=1\linewidth]{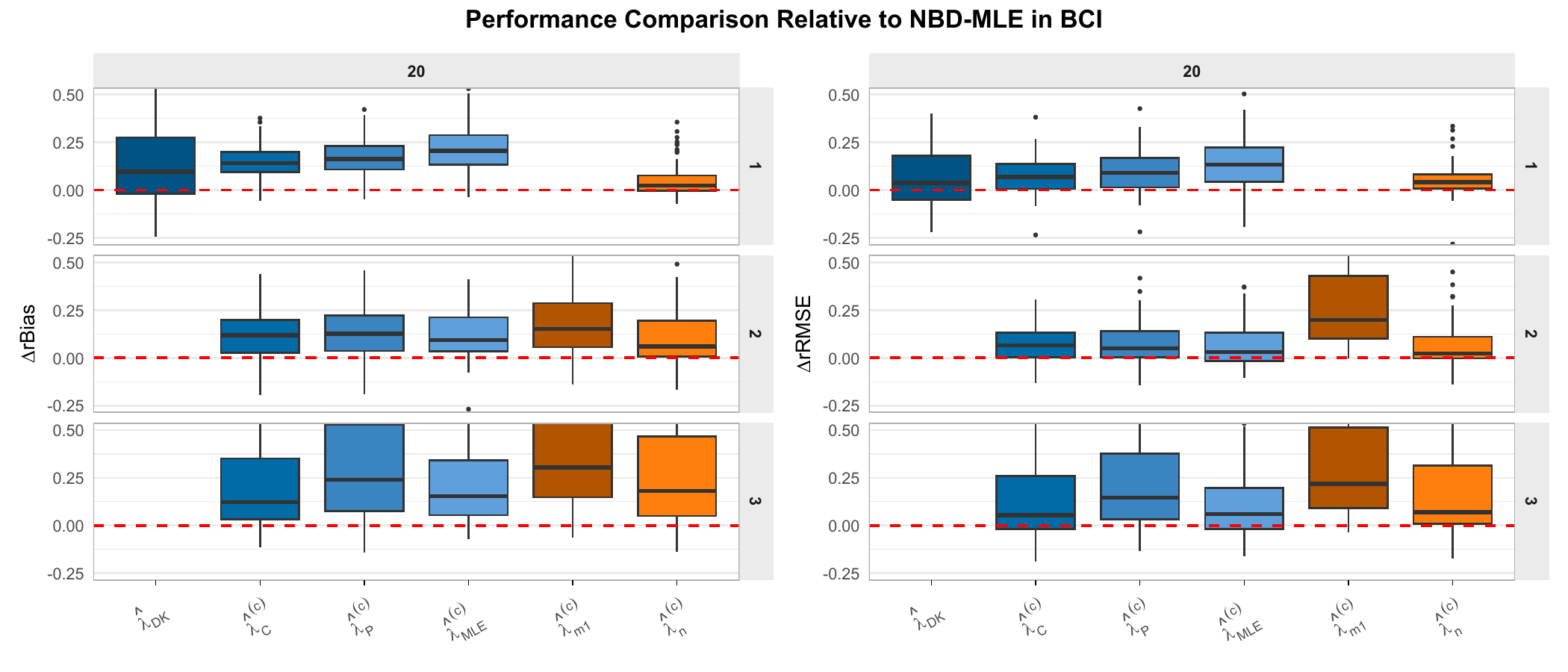}
		\caption{
			Differences in absolute relative bias ($|\text{rBias}|$) and relative root mean squared error (rRMSE) between six censored density estimators and the censored NBD maximum likelihood estimator, $\hat{\lambda}_{\text{n,MLE}}^{(c)}$, for the BCI plot with $C=20$\,m and $\ell=1,2,3$.
			Left panel: $|\text{rBias}|_{\text{method}}-|\text{rBias}|_{\text{NBD-MLE}}$.
			Right panel: $\text{rRMSE}_{\text{method}}-\text{rRMSE}_{\text{NBD-MLE}}$.
			For each species and each nearest-neighbour order, rBias and rRMSE were computed from 200 independent sampling replicates; boxplots summarise the resulting across-species distributions for species with at least 500 individuals.
			The red dashed line at zero indicates equal performance under the corresponding metric; values above zero indicate larger error than $\hat{\lambda}_{\text{n,MLE}}^{(c)}$.
			The estimator $\hat{\lambda}_{\text{DK}}$ is applicable only for $\ell=1$ and the estimator $\hat{\lambda}_{m1}^{(c)}$ is not applicable for $\ell=1$.
		}
		\label{fig:bci_performance}
	\end{figure}

	The empirical analyses confirmed the patterns observed in the simulation study, with the NBD-based maximum likelihood estimator $\hat{\lambda}_{\text{n,MLE}}^{(c)}$ consistently outperforming all other methods.
	Figure~\ref{fig:bci_performance} summarises the performance of seven censored estimators applied to 112 tree species (abundance $\geq 500$ individuals) in the BCI 50‑ha plot, using a fixed maximum search radius $C = 20$\,m and nearest‑neighbour orders $\ell = 1,2,3$.
	Because $\hat{\lambda}_{\text{n,MLE}}^{(c)}$ was uniformly the best estimator, we display the results as differences in absolute relative bias ($|\text{rBias}|$) and relative root mean squared error ($\text{rRMSE}$) between each method and $\hat{\lambda}_{\text{n,MLE}}^{(c)}$.
	Positive values indicate that a method yields a larger error than $\hat{\lambda}_{\text{n,MLE}}^{(c)}$.
	
	Across all search radii and nearest‑neighbour orders, $\hat{\lambda}_{\text{n,MLE}}^{(c)}$ usually achieved the smallest median $|\text{rBias}|$ and the smallest median $\text{rRMSE}$ among all estimators. In particular, compared with the widely used existing corrections $\hat{\lambda}_{\text{DK}}$ and $\hat{\lambda}_{C}^{(c)}$, the proposed $\hat{\lambda}_{\text{n,MLE}}^{(c)}$ reduced the median absolute relative bias by approximately $0.10$, corresponding to a relative improvement over 30\%. These findings were broadly consistent across different censoring radii ($C = 10, 20, 30, 40$\,m) and were reproduced in the Harvard Forest 35‑ha plot (see Supplement Section~\S4). 
	Tables S1 and S2 in Supplement Section~\S4 summarise the across‑species distributions of these improvements. The advantage of $\hat{\lambda}_{\text{n,MLE}}^{(c)}$ was most pronounced at small search radii: for $C = 10$\,m, the median relative improvement exceeded 60\% for both $\hat{\lambda}_{\text{DK}}$ and $\hat{\lambda}_{C}^{(c)}$ in both forest plots, reaching 88\% for $\hat{\lambda}_{\text{DK}}$ at BCI (Table S2). Supplement Section~\S4 also provides boxplots of raw $\text{rBias}$ and $\text{rRMSE}$, as well as the corresponding absolute and relative differences with respect to $\hat{\lambda}_{\text{n,MLE}}^{(c)}$, for all methods across all combinations of $C$ and $\ell$ in both the BCI and Harvard Forest plots.
	
\begin{figure}[!t]
		\centering
		\includegraphics[width=1\linewidth]{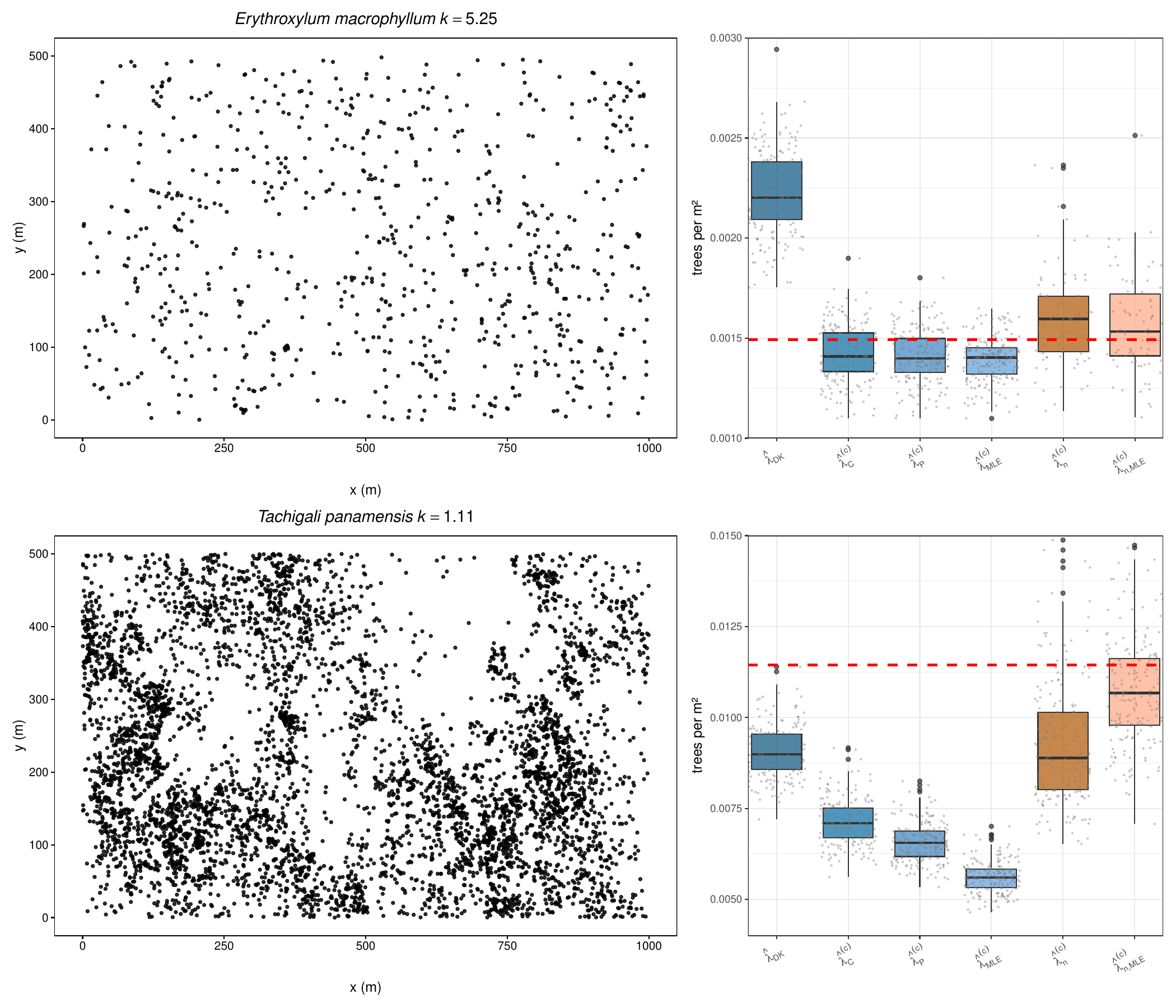}
		\caption{Spatial distributions and estimator performance for two BCI species. \textit{Top row}: Spatial point pattern of \textit{Erythroxylum macrophyllum} (left) and corresponding boxplots of 200 replicate density estimates (right). The aggregation parameter $k$ is estimated by MLE under a negative binomial model using 25m $\times$ 25m gridded census data. \textit{Bottom row}: Spatial point pattern of \textit{Tachigali versicolor} (left) and corresponding boxplots (right). For both species, the sampling design uses $n=120$ focal points, $\ell=1$ nearest neighbour, and a maximum search radius of $C=20$m. Red dashed lines indicate the true population densities. Since $\ell=1$, $\hat{\lambda}_{m1}^{(c)}$ is not applicable and is therefore omitted.}
		\label{fig:species_example}
	\end{figure}

	For a direct visual comparison, Figure~\ref{fig:species_example} illustrates the density estimates for two representative species with contrasting spatial structures. The top panels show \textit{Erythroxylum macrophyllum}, whose distribution is close to complete spatial randomness. In this case, all estimators are nearly unbiased, except for $\hat{\lambda}_{\text{DK}}$, which exhibits a positive bias consistent with the CSR simulation results in Table~\ref{tab:csr_sim_results}. The bottom panels show \textit{Tachigali versicolor}, a strongly aggregated species. Here, all estimators underestimate the true density, but $\hat{\lambda}_{\text{n,MLE}}^{(c)}$ produces the smallest negative bias.
	
	\section{Discussion}
	
	This study develops a unified framework for estimating population density from point-centred quarter method (PCQM) data under right-censoring. By extending both moment-based and likelihood-based estimators under the Poisson model for complete spatial randomness and the negative binomial distribution (NBD) model for aggregated populations, we provide a coherent set of tools for analysing censored distance data under realistic field constraints. Across simulations and empirical forest plot analyses, the NBD-based maximum likelihood estimator for censored data, $\hat{\lambda}_{\text{n,MLE}}^{(c)}$, showed the strongest overall performance when spatial aggregation was present.
	
	From the perspective of practical application, our unified framework also simplifies the analytical workflow for censored PCQM data.
	Existing corrections for right-censored distances, such as those of \citet{warde1981correction} and \citet{dahdouh2006empirical}, are valid only under the assumption of complete spatial randomness. Consequently, a practitioner using these methods must first test whether the observed spatial pattern is consistent with CSR, for instance by employing the test statistic proposed by \citet{eberhardt1967developments}.
	If the CSR null hypothesis is rejected, the available censored-data estimators are no longer justified, leaving no principled alternative.
	Our NBD-based framework removes this obstacle: because it accommodates both random and aggregated spatial structures within a single likelihood, $\hat{\lambda}_{\text{n,MLE}}^{(c)}$ can be applied directly without a preliminary test of spatial randomness.
	
	Although the NBD-based likelihood estimator $\hat{\lambda}_{\text{n,MLE}}^{(c)}$ delivers the most accurate density estimates under a wide range of spatial structures, its use requires numerical optimization and does not yield a closed-form solution. By contrast, the Poisson-based censored estimators $\hat{\lambda}_{C}^{(c)}$ and $\hat{\lambda}_{P}^{(c)}$ rely only on a Gamma quantile evaluation, and the censored Poisson MLE $\hat{\lambda}_{\text{MLE}}^{(c)}$ has an explicit closed form for the commonly used case $\ell=1$ and requires only a univariate numerical optimization for $\ell \geq 2$. All are computationally trivial compared with the bivariate optimization required by the NBD-based MLE. They therefore provide useful baseline estimates and may be preferred in exploratory analyses or when computational simplicity is paramount.
	
	One practical implication of our results concerns the trade-off between the maximum search radius ($C$) and the nearest-neighbour order ($\ell$) in censored sampling designs. In uncensored settings, increasing $\ell$ can improve estimation efficiency \citep{Kronenfeld2009APD,khan2016evaluation}. Under a finite search radius, however, this advantage is substantially reduced because the probability of censoring increases rapidly with $\ell$ (Supplement Section \S4). As a result, designs based on lower neighbour orders, especially $\ell=1$ or $2$, can provide a more reliable balance between information gain and censoring-induced bias when truncation is substantial.
	
	Despite these practical insights, several limitations and avenues for future methodological extension remain. First, although aggregation is common, the NBD framework does not capture all ecologically relevant spatial patterns, particularly strongly inhibited or highly regular structures \citep{cogbill2018retrospective}. Second, the moment-based censored NBD estimators, $\hat{\lambda}_{n}^{(c)}$ and $\hat{\lambda}_{m1}^{(c)}$, require an initial density estimate and therefore involve an additional approximation step. Although our results suggest that this approximation is acceptable in the settings considered here, it may become less reliable under extreme censoring or in populations whose spatial structure departs substantially from the NBD assumption. Future work could therefore consider extensions to alternative point-process models and investigate adaptive strategies for selecting search radius and neighbour order jointly.
	
	Overall, our results show that right-censoring and spatial aggregation need to be addressed jointly when applying plotless distance-based sampling in ecological surveys. By bringing these two features into a common inferential framework, this study broadens the practical scope of PCQM-based density estimation and provides a rigorously validated basis for analysing censored point-to-tree distance data under realistic field conditions.
	
	
	\section*{Acknowledgements}
	
	D.X. was funded by the National Nature Science Foundation of China (32471623) and the Innovation Program of Shanghai Municipal Education Commission (2023ZKZD36). We thank the Forest Global Earth Observatory (ForestGEO) network for making the BCI and HF data publicly available.
	
	\bibliographystyle{chicago}
	\bibliography{refs}

@article{cogbill2023surveyor,
  title={Surveyor and Analyst Biases in Forest Density Estimation from {U}nited {S}tates Public Land Surveys},
  author={Cogbill, Charles V.},
  journal={Ecosphere},
  volume={14},
  number={8},
  pages={e4647},
  year={2023},
  doi={10.1002/ecs2.4647}
}

@article{santini2024tetradensity,
  title={Tetra{DENSITY} 2.0—A Database of Population Density Estimates in Tetrapods},
  author={Santini, Luca and Mendez Angarita, V. Y. and Karoulis, C. and Fundarò, D. and Pranzini, N. and Vivaldi, C. and Zhang, T. and Zampetti, A. and Gargano, S. J. and Mirante, D. and Paltrinieri, L.},
  journal={Global Ecology and Biogeography},
  volume={33},
  number={5},
  pages={e13929},
  year={2024},
  doi={10.1111/geb.13929}
}

@article{levine2017evaluating,
  title={Evaluating a new method for reconstructing forest conditions from {G}eneral {L}and {O}ffice survey records},
  author={Levine, Carrie R. and Cogbill, Charles V. and Collins, Brandon M. and Larson, Andrew J. and Lutz, James A. and North, Malcolm P. and Restaino, Christina M. and Safford, Hugh D. and Stephens, Scott L. and Battles, John J.},
  journal={Ecological Applications},
  volume={27},
  number={5},
  pages={1498--1513},
  year={2017},
  doi={10.1002/eap.1543}
}

@article{cogbill2018retrospective,
  title={A retrospective on the accuracy and precision of plotless forest density estimators in ecological studies},
  author={Cogbill, Charles V and Thurman, Amanda L and Williams, John W and Mladenoff, David J and Goring, Simon J},
  journal={Ecosphere},
  volume={9},
  number={2},
  pages={e02187},
  year={2018},
  doi={10.1002/ecs2.2187}
}

@article{cottam1953some,
  title={Some sampling characteristics of a population of randomly dispersed individuals},
  author={Cottam, Grant and Curtis, JT and Hale, BW},
  journal={Ecology},
  volume={34},
  number={4},
  pages={741--757},
  year={1953},
  doi={10.2307/1931337}
}

@article{cottam1956use,
  title={The use of distance measures in phytosociological sampling},
  author={Cottam, Grant and Curtis, JT},
  journal={Ecology},
  volume={37},
  number={3},
  pages={451--460},
  year={1956},
  doi={10.2307/1930167}
}

@article{dahdouh2006empirical,
  title={Empirical estimate of the reliability of the use of the point-centred quarter method ({PCQM}): Solutions to ambiguous field situations and description of the {PCQM}+ protocol},
  author={Dahdouh-Guebas, Farid and Koedam, Nico},
  journal={Forest Ecology and Management},
  volume={228},
  number={1-3},
  pages={1--18},
  year={2006},
  doi={10.1016/j.foreco.2006.02.021}
}

@Book{mitchell2023quantitative,
  author        = {Kevin Mitchell},
  title         = {Quantitative Analysis by the Point-Centered Quarter Method},
  year          = {2023},
  note          = {arXiv: 1010.3303},
  archiveprefix = {arXiv},
  eprint        = {1010.3303},
  primaryclass  = {q-bio.QM},
  url           = {https://arxiv.org/abs/1010.3303},
}

@article{morisita1954estimation,
  title={Estimation of population density by spacing method},
  author={Morisita, Masaaki},
  journal={Memoirs of the Faculty of Science, Kyushu University. Series E, Biology},
  volume={1},
  pages={187--197},
  year={1954}
}

@article{morisita1957new,
  title={A new method for the estimation of density by the spacing method applicable to nonrandomly distributed populations},
  author={Morisita, Masaaki},
  journal={Physiology and Ecology},
  volume={7},
  pages={134--144},
  year={1957}
}

@article{pollard1971distance,
  title={On distance estimators of density in randomly distributed forests},
  author={Pollard, JH},
  journal={Biometrics},
  volume={27},
  number={4},
  pages={991--1002},
  year={1971},
  doi={10.2307/2528833}
}

@article{shen2020distance,
  author = {Shen, Guochun and Wang, Xihua and He, Fangliang},
  year = {2020},
  title = {Distance‐based methods for estimating density of nonrandomly distributed populations},
  volume = {101},
  number = {10},
  journal = {Ecology},
  pages = {e03143},
  doi = {10.1002/ecy.3143}
}

@article{warde1981correction,
  title={A correction factor table for missing point-center quarter data},
  author={Warde, William and Petranka, James W},
  journal={Ecology},
  volume={62},
  number={2},
  pages={491--494},
  year={1981},
  doi={10.2307/1936723}
}

@article{he1997distribution,
  title={Distribution patterns of tree species in a Malaysian tropical rain forest},
  author={He, Fangliang and Legendre, Pierre and LaFrankie, James V},
  journal={Journal of Vegetation Science},
  volume={8},
  number={1},
  pages={105--114},
  year={1997},
  doi={10.2307/3237250}
}

@article{eberhardt1967developments,
  author = {Eberhardt, L. L.},
  title = {Some Developments in `Distance Sampling'},
  journal = {Biometrics},
  year = {1967},
  volume = {23},
  number = {2},
  pages = {207--216},
  publisher = {International Biometric Society},
  url = {https://www.jstor.org/stable/2528156}
}

@article{gao2013detecting,
  author = {Gao, M.},
  title = {Detecting Spatial Aggregation from Distance Sampling: A Probability Distribution Model of Nearest Neighbor Distance},
  journal = {Ecological Research},
  year = {2013},
  volume = {28},
  number = {3},
  pages = {397--405},
  month = {may},
  doi = {10.1007/s11284-013-1029-x}
}

@Article{Stoklosa2022NBD,
AUTHOR = {Stoklosa, Jakub and Blakey, Rachel V. and Hui, Francis K. C.},
TITLE = {An Overview of Modern Applications of Negative Binomial Modelling in Ecology and Biodiversity},
JOURNAL = {Diversity},
VOLUME = {14},
YEAR = {2022},
NUMBER = {5},
pages = {320},
URL = {https://www.mdpi.com/1424-2818/14/5/320},
ISSN = {1424-2818},
DOI = {10.3390/d14050320}
}

@article{Thomas1949,
 ISSN = {00063444, 14643510},
 URL = {http://www.jstor.org/stable/2332526},
 author = {Marjorie Thomas},
 journal = {Biometrika},
 number = {1/2},
 pages = {18--25},
 publisher = {[Oxford University Press, Biometrika Trust]},
 title = {A Generalization of Poisson's Binomial Limit For use in Ecology},
 urldate = {2025-12-16},
 volume = {36},
 year = {1949}
}

@article{OLSSON2014Latin,
title = {On Latin hypercube sampling for structural reliability analysis},
journal = {Structural Safety},
volume = {25},
number = {1},
pages = {47-68},
year = {2003},
issn = {0167-4730},
doi = {https://doi.org/10.1016/S0167-4730(02)00039-5},
url = {https://www.sciencedirect.com/science/article/pii/S0167473002000395},
author = {A. Olsson and G. Sandberg and O. Dahlblom}
}

@article{ZILLIO2010modelling,
author = {Zillio, Tommaso and He, Fangliang},
title = {Modeling spatial aggregation of finite populations},
journal = {Ecology},
volume = {91},
number = {12},
pages = {3698-3706},
doi = {https://doi.org/10.1890/09-2233.1},
url = {https://esajournals.onlinelibrary.wiley.com/doi/abs/10.1890/09-2233.1},
eprint = {https://esajournals.onlinelibrary.wiley.com/doi/pdf/10.1890/09-2233.1},
year = {2010}
}

@article{Condit2018community,
author = {Chen, Youhua and Shen, Tsung-Jen and Condit, Richard and Hubbell, Stephen P.},
title = {Community-level species’ correlated distribution can be scale-independent and related to the evenness of abundance},
journal = {Ecology},
volume = {99},
number = {12},
pages = {2787-2800},
doi = {https://doi.org/10.1002/ecy.2544},
url = {https://esajournals.onlinelibrary.wiley.com/doi/abs/10.1002/ecy.2544},
eprint = {https://esajournals.onlinelibrary.wiley.com/doi/pdf/10.1002/ecy.2544},
year = {2018}
}

@article{Khan2016Evaluation,
  author = {Khan, Md Nabiul Islam and Hijbeek, Renske and Berger, Uta and Koedam, Nico and Grueters, Uwe and Islam, S. M. Zahirul and Hasan, Md Asadul and Dahdouh-Guebas, Farid},
  title = {An Evaluation of the Plant Density Estimator the Point-Centred Quarter Method (PCQM) Using Monte Carlo Simulation},
  journal = {PLOS ONE},
  volume = {11},
  number = {6},
  pages = {e0157985},
  year = {2016},
  doi = {10.1371/journal.pone.0157985},
  issn = {1932-6203}
}

@article{Kronenfeld2009APD,
  title={A Plotless Density Estimator Based on the Asymptotic Limit of Ordered Distance Estimation Values},
  author={Kronenfeld, Barry J.},
  journal={Forest Science},
  volume={55},
  number={4},
  pages={283--292},
  year={2009},
  doi={10.1093/forestscience/55.4.283},
  url={https://doi.org/10.1093/forestscience/55.4.283}
}

@Manual{carnell2026lhs,
  title = {lhs: Latin Hypercube Samples},
  author = {Rob Carnell},
  year = {2026},
  note = {R package version 1.2.1},
  url = {https://github.com/bertcarnell/lhs},
}

@book{baddeley2015spatstat,
  title={Spatial Point Patterns: Methodology and Applications with R},
  author={Baddeley, Adrian and Rubak, Ege and Turner, Rolf},
  year={2015},
  publisher={Chapman and Hall/CRC},
  address={Boca Raton, FL}
}

@Article{notarangelo2023performance,
  author  = {Notarangelo, Monica and Carrer, Marco and Lingua, Emanuele and Puletti, Nicola and Torresan, Chiara},
  journal = {iForest -- Biogeosciences and Forestry},
  title   = {Performance assessment of two plotless sampling methods for density estimation applied to some Alpine forests of northeastern Italy},
  year    = {2023},
  number  = {6},
  pages   = {385--391},
  volume  = {16},
  doi     = {10.3832/ifor4335-016},
}

@Article{pommerening2024distance,
  author  = {Pommerening, Arne and Sterba, Hubert and Eskelson, Bianca N. I.},
  journal = {Ecological Indicators},
  title   = {Distance and {T}-square sampling for spatial measures of tree diversity},
  year    = {2024},
  pages   = {111995},
  volume  = {163},
  doi     = {10.1016/j.ecolind.2024.111995},
}

@Article{mienna2024quantifying,
  author  = {Mienna, Ida Marielle and Klanderud, Kari and N{\ae}sset, Erik and Gobakken, Terje and Bollands{\aa}s, Ole Martin},
  journal = {Ecosphere},
  title   = {Quantifying the roles of climate, herbivory, topography, and vegetation on tree establishment in the treeline ecotone},
  year    = {2024},
  number  = {4},
  pages   = {e4845},
  volume  = {15},
  doi     = {10.1002/ecs2.4845},
}

@article{bryant2005forest,
  title={Forest community analysis and the point-centered quarter method},
  author={Bryant, David M. and Ducey, Mark J. and Innes, James C. and Lee, Thomas D. and Eckert, Robert T. and Zarin, Daniel Jacob},
  journal={Plant Ecology},
  volume={175},
  number={2},
  pages={193--203},
  year={2005},
  doi={10.1007/s11258-005-0013-0}
}

@article{wiegand2025latitudinal,
  title={Latitudinal scaling of aggregation with abundance and coexistence in forests},
  author={Wiegand, Thorsten and Wang, Xugao and Fischer, Samuel M. and Kraft, Nathan J. B. and Bourg, Norman A. and Brockelman, Warren Y. and Cao, Guanghong and Cao, Min and Chanthorn, Wirong and Chu, Chengjin and Davies, Stuart and Ediriweera, Sisira and Gunatilleke, C. V. Savitri and Gunatilleke, I. A. U. Nimal and Hao, Zhanqing and Howe, Robert and Jiang, Mingxi and Jin, Guangze and Kress, W. John and Li, Buhang and Lian, Juyu and Lin, Luxiang and Liu, Feng and Ma, Keping and McShea, William and Mi, Xiangcheng and Myers, Jonathan A. and Nathalang, Anuttara and Orwig, David A. and Shen, Guochun and Su, Sheng-Hsin and Sun, I-Fang and Wang, Xihua and Wolf, Amy and Yan, Enrong and Ye, Wanhui and Zhu, Yan and Huth, Andreas},
  journal={Nature},
  volume={640},
  number={8060},
  pages={967-973},
  year={2025},
  doi={10.1038/s41586-025-08604-z}
}

@article{condit2019complete,
  title={Complete data from the {Barro} {Colorado} 50-ha plot: 423617 trees, 35 years},
  author={Condit, Richard and P{\'e}rez, Rolando and Aguilar, Salom{\'o}n and Lao, Suzanne and Foster, Robin and Hubbell, Stephen P},
  year={2019},
  publisher={ForestGEO Data Portal}
}

@article{orwig2022harvard,
  title={{Harvard} {Forest} {CTFS}-{ForestGEO} Mapped Forest Plot since 2014. {Harvard} {Forest} {Data} {Archive}: {HF}253.5},
  author={Orwig, David and Foster, David and Ellison, Aaron},
  year={2022},
  publisher={ForestGEO Data Portal}
}
	
	\pagebreak
	\appendix
	\renewcommand{\thesection}{S\arabic{section}}
	\renewcommand{\thesubsection}{S\arabic{section}.\arabic{subsection}}
	\setcounter{figure}{0}
	\setcounter{table}{0}
	\setcounter{equation}{0}
	\renewcommand{\thefigure}{S\arabic{figure}}
	\renewcommand{\thetable}{S\arabic{table}}
	\renewcommand{\theequation}{S\arabic{equation}}
	
	\section*{Supporting Information}
	
	\noindent Huang, W., G. Shen, D. Xing, and J. Zhao. 2026. A Unified Framework for Density Estimation under Right-Censored Point-Centred Quarter Sampling.
	
	\section{Moment Adjustment for Right-Censored Data under Poisson case}\label{sec:moment_adjustment}
	
	This section provides a detailed derivation of the generalized moment-based density estimators for right-censored distance sampling data, extending the original work of \citet{warde1981correction} from the special case of $\ell=1$ to the general case of the $\ell$th nearest neighbor in $q$ sectors and to other kinds of moments-based estimators.
	
	Consider a sampling design with $q$ equal-angle sectors per focal point. Under the assumption of complete spatial randomness (CSR) with density $\lambda$ (individuals per unit area), the number of individuals $m$ in a sector of radius $r$ follows a Poisson distribution with parameter:
	\[
	m = \lambda \pi r^2 / q.
	\]
	The cumulative distribution function of the distance $R$ to the $\ell$th nearest neighbor within a sector is
	\[
	\prob(R \leq r) = 1 - \sum_{i=0}^{\ell-1} \frac{(\lambda \pi r^2/q)^i}{i!} e^{-\lambda \pi r^2/q}.
	\]
	The probability density function is
	\[
	f(r) = \frac{2\pi \lambda r}{q} \frac{(\lambda \pi r^2/q)^{\ell-1}}{(\ell-1)!} e^{-\lambda \pi r^2/q}.
	\]
	The $u$th moment of $R$ under complete sampling is given by
	\begin{equation}\label{eq:ERu}
		\bE[R^u] = \int_0^\infty r^u f(r) dr = \left( \frac{q}{\pi \lambda} \right)^{u/2} \frac{\Gamma\left(\ell + \frac{u}{2}\right)}{(\ell-1)!}.
	\end{equation}
	In truncated sampling with maximum search radius $C$, we only observe distances when $R \leq C$. The probability of observing at least $\ell$ individuals within radius $C$ is
	\[
	P = 1 - \sum_{i=0}^{\ell-1} \frac{m_C^i}{i!} e^{-m_C}, \quad \text{where } m_C = \lambda \pi C^2/q.
	\]
	Practically, $P$ is estimated by the observed proportion of non-censored sectors: $\hat{P} = 1 - n_0/(nq)$.
	The conditional $u$th moment given $R \leq C$ is
	\[
	\bE[R^u \mid R \leq C] = \frac{\int_0^C r^u f(r) dr}{P}.
	\]
	The numerator is
	\[
	\int_0^C r^u f(r) dr = \frac{2\pi \lambda}{q} \frac{(\lambda \pi/q)^{\ell-1}}{(\ell-1)!} \int_0^C r^{2\ell + u - 1} e^{-b r^2} dr,
	\]
	where $b = \lambda \pi/q$.
	
	Using the substitution $t = b r^2$ and the definition of the lower incomplete gamma function $\gamma(s, x) = \int_0^x t^{s-1} e^{-t} dt$, we obtain
	\[
	\int_0^C r^{2\ell + u - 1} e^{-b r^2} dr = \frac{1}{2} b^{-(\ell + u/2)} \gamma\left(\ell + \frac{u}{2}, m_C\right).
	\]
	Therefore,
	\[
	\int_0^C r^u f(r) dr = \frac{\gamma\left(\ell + \frac{u}{2}, m_C\right)}{(\ell-1)!} \left( \frac{q}{\pi \lambda} \right)^{u/2}.
	\]
	The conditional moment becomes
	\begin{equation}
		\bE[R^u \mid R \leq C] = \frac{\gamma\left(\ell + \frac{u}{2}, m_C\right)}{(\ell-1)! P} \left( \frac{q}{\pi \lambda} \right)^{u/2}.
	\end{equation}
	
	Comparing with the unconditional moment $\bE[R^u]$ \eqref{eq:ERu}, we establish the relationship:
	\[
	\bE[R^u] = \bE[R^u \mid R \leq C] \cdot \frac{\Gamma\left(\ell + \frac{u}{2}\right)}{\gamma\left(\ell + \frac{u}{2}, m_C\right)} \cdot P.
	\]
	
	In practice, we compute the sample conditional moment $\hat{M_u} = \frac{1}{nq-n_0} \sum_{t=1}^{nq-n_0} r_t^{'u}$ from the $nq-n_0$ non-censored observations. The adjusted unconditional moment estimator is therefore
	\[
	{M}_u = \hat{M_u} \cdot \frac{\Gamma\left(\ell + \frac{u}{2}\right)}{\gamma\left(\ell + \frac{u}{2}, \hat{m_C}\right)} \cdot \hat{P},
	\]
	where $\hat{m_C}$ is the solution to $\frac{\gamma(\ell, \hat{m_C})}{\Gamma(\ell)} = \hat{P}$, providing an estimate of $m_C$, since
	\[
	\frac{\gamma(\ell, x)}{\Gamma(\ell)} = 1 - \sum_{k=0}^{\ell-1} \frac{x^k}{k!} e^{-x}.
	\]

	\section{Theoretical Justification for Moment Adjustment under NBD Model}\label{sec:nbd_adjustment}
	
	\subsection{Why is the Adjusted Moment \(\hat{\bE[R^u]}\) Justified?}
	For a sampling design with \( n \) focal points, each divided into \( q \) equal‑angle sectors, we measure distances \( R_j \) (\( j = 1, \dots, nq \)) to the \( \ell \)th nearest individual. A maximum search radius \( C \) results in right‑censored observations (\( R_j > C \)). The adjusted \( u \)th sample moment is
	\[
	\hat{\bE[R^u]} \approx \frac{1}{nq} \sum_{j=1}^{nq} \left[ \mathbbm{1}_{\{R_j \leq C\}} R_j^u + \mathbbm{1}_{\{R_j > C\}} \bE[R^u \mid R > C; k = +\infty, \lambda = \lambda_{\text{init}}] \right],
	\]
	where \( \mathbbm{1}_{\{\cdot\}} \) is the indicator function, and \( \bE[R^u \mid R > C; k = +\infty, \lambda = \lambda_{\text{init}}] \) is the conditional expectation under the CSR assumption (\( k = +\infty \)) with an initial density estimate \( \lambda_{\text{init}} \), typically from traditional methods \citet{warde1981correction,dahdouh2006empirical} or our generalized estimator. The CSR conditional expectation is
	\[
	\bE[R^u \mid R > C; k = +\infty, \lambda = \lambda_{\text{init}}] = \left( \pi \lambda_{\text{init}} q^{-1} \right)^{-u/2} \frac{\Gamma\left( \ell + \frac{u}{2}, \pi \lambda_{\text{init}} q^{-1} C^2 \right)}{\Gamma\left( \ell, \pi \lambda_{\text{init}} q^{-1} C^2 \right)},
	\]
	where \( \Gamma(a, x) \) is the upper incomplete gamma function $\Gamma(a, x) = \int_x^{\infty} t^{a-1} e^{-t} \, dt$.
	
	This formulation is justified by the Strong Law of Large Numbers (SLLN). If the true parameters \( k \), \( \lambda \) were known, we define
	\[
	Z_j = \mathbbm{1}_{\{R_j \leq C\}} R_j^u + \mathbbm{1}_{\{R_j > C\}} \bE[R^u \mid R > C; k, \lambda],
	\]
	with \( \bE[Z_j] = \bE[R^u] \). By the SLLN, \( \frac{1}{nq} \sum_{j=1}^{nq} Z_j \to \bE[R^u] \). Since the true parameters are unknown, we approximate them with \( k = +\infty \) and \( \lambda_{\text{init}} \). The bias can be estimated by
	\[
	\begin{aligned}
		\text{Bias} &\approx \prob(R > C) \cdot \left[ \bE[R^u \mid R > C; k = +\infty, \lambda = \lambda_{\text{init}}] - \bE[R^u \mid R > C; k, \lambda] \right] \\
		&\approx \prob(R > C) \cdot \left[ \bE[R^u \mid R > C; k = +\infty, \lambda = \lambda] - \bE[R^u \mid R > C; k, \lambda] \right]
	\end{aligned}
	\]
	where
	\[
	\prob(R > C) = 1 - I_w(\ell, k),
	\]
	and
	\[
	\bE[R^u \mid R > C; k, \lambda] = \left( \frac{k q}{\pi \lambda} \right)^{u/2} \frac{\Gamma\left(\ell + \frac{u}{2}\right) \Gamma\left(k - \frac{u}{2}\right)}{\Gamma(\ell) \Gamma(k)} \cdot \frac{1 - I_w\left( \ell + \frac{u}{2}, k - \frac{u}{2} \right)}{1 - I_w(\ell, k)},
	\]
	with \( a = \pi \lambda q^{-1} \), \( u_C = \frac{a C^2}{k} \), \( w = \frac{u_C}{1 + u_C} \), and \( I_w(a, b) \) is the regularized incomplete Beta function.
	
	When \( k \) is small (i.e., approaching the theoretical lower bound of \( k > \frac{u}{2} \) from above) and \( C \) is relatively large, the probability \( \prob(R > C) = 1 - I_w(\ell, k) \) tends to 0. This is because \( u_C = \frac{a C^2}{k} \) becomes large, causing \( w = \frac{u_C}{1 + u_C} \) to approach 1, and thus \( I_w(\ell, k) \to 1 \). Consequently, the bias term is dominated by \( \prob(R > C) \) and becomes negligible.
	
	When \( k \) is large, the difference in conditional expectations, \( \bE[R^u \mid R > C; k = +\infty, \lambda = \lambda] - \bE[R^u \mid R > C; k, \lambda] \), is of order \( O(1/k) \). This can be rigorously shown via a Taylor expansion: for general \( \ell \) and \( u \) (with \( k > \frac{u}{2} \) to ensure the moments exist), the conditional expectation under NBD can be expanded as
	\[
	\bE[R^u \mid R > C; k, \lambda] = \bE[R^u \mid R > C; k = +\infty, \lambda = \lambda] + \frac{\Delta_1}{k} + O\left(\frac{1}{k^2}\right),
	\]
	where \( \bE[R^u \mid R > C; k = +\infty, \lambda = \lambda] = a^{-u/2} \frac{\Gamma(\alpha, T)}{\Gamma(\ell, T)} \) with \( a = \pi \lambda q^{-1} \), \( \alpha = \ell + \frac{u}{2} \), and \( T = a C^2 \), and \( \Delta_1 \) is an explicit constant given by
	\[
	\Delta_1 = a^{-u/2} \frac{ -\ell \Gamma(\alpha+1, T) \Gamma(\ell, T) + \frac{1}{2} \Gamma(\alpha+2, T) \Gamma(\ell, T) + \ell \Gamma(\alpha, T) \Gamma(\ell+1, T) - \frac{1}{2} \Gamma(\alpha, T) \Gamma(\ell+2, T) }{ \Gamma(\ell, T)^2 }.
	\]
	Thus, the difference between the CSR-based approximation and the true NBD conditional expectation vanishes as \( k \) increases, confirming the robustness of our moment adjustment for moderately aggregated or spatially random populations.
	
	Conversely, for a fixed maximum search radius \( C \), as \( k \to 0^+ \) (indicating extremely strong spatial aggregation of the population), the censoring probability \( \prob(R > C) = 1 - I_w(\ell, k) \) tends to 1. In this scenario, individuals are so highly clustered that the vast majority of sampling sectors contain no individuals within the predefined search radius \( C \). While the moment adjustment formula remains mathematically well-defined, the practical utility of this moment-based approach is severely limited: the sample contains almost no effective distance information, leading to highly unstable and unreliable estimates of population density. This highlights a key practical constraint: the search radius \( C \) must be appropriately scaled to match the spatial aggregation scale of the target population to ensure sufficient non-censored observations for valid inference.
	
	\subsection{Why is the Adjusted Moment \(\hat{\bE[R^u]}\) Better than \(M_u\)?}
	We characterize the asymptotic behavior of \(\hat{\bE[R^u]}\) and \(M_u\) as the total number of sampled sectors \(N = nq \to \infty\) (i.e., the sample size \(n\) of focal points tends to infinity).
	
	\paragraph{Asymptotic Limit of \(M_u\).}  
	When \(N \to \infty\), the empirical censoring proportion \(p_0^{\text{obs}} = n_0/N\) converges to the theoretical censoring probability \(p_0 = 1 - \prob(R \leq C; k, \lambda)\), where \(\prob(R \leq C; k, \lambda) = I_w(\ell, k)\). The Poisson-based adjusted moment \(M_u\) converges to its asymptotic limit:
	\[
	M_u^\infty = \bE[R^u \mid R \leq C; k, \lambda] \cdot \frac{\Gamma\left(\ell + \frac{u}{2}\right)}{\gamma\left(\ell + \frac{u}{2}, \hat{m}_C^\infty\right)} \cdot (1 - p_0),
	\]
	where \(\hat{m}_C^\infty \) is the solution to \(\gamma(\ell, \hat{m}_C^\infty)/\Gamma(\ell) = 1 - p_0\) and \(\bE[R^u \mid R \leq C; k, \lambda]\) is the theoretical conditional expectation of \(R^u\) given \(R \leq C\) under the NBD model.
	
	\paragraph{Asymptotic Limit of \(\hat{\bE[R^u]}\).}  
	For the censored-adjusted moment \(\hat{\bE[R^u]}\) (tailored for the NBD model), as \(N \to \infty\), the initial density estimate \(\lambda_{\text{init}}\) (derived from the Pollard-type censored estimator) converges to its theoretical value \(\lambda_{\text{init}}^\infty = \frac{\ell q}{\pi M_2^\infty}\). Consequently, \(\hat{\bE[R^u]}\) converges to
	\[
	\hat{\bE[R^u]}^\infty = (1 - p_0) \cdot \bE[R^u \mid R \leq C; k, \lambda] + p_0 \cdot \bE[R^u \mid R > C; k = \infty, \lambda_{\text{init}}^\infty],
	\]
	where \(\bE[R^u \mid R > C; k = \infty, \lambda_{\text{init}}^\infty]\) denotes the theoretical conditional expectation of \(R^u\) given \(R > C\) under CSR with density \(\lambda_{\text{init}}^\infty\). 
	
	\paragraph{Asymptotic Bias Comparison.}  
	Let \(\mu_u^* = \bE[R^u]\) be the true \(u\)-th moment of \(R\) under the uncensored NBD model. The asymptotic bias of \(M_u\) is defined as \(B(M_u) = M_u^\infty - \mu_u^*\), and the asymptotic bias of \(\hat{\bE[R^u]}\) is \(B(\hat{\bE[R^u]}) = \hat{\bE[R^u]}^\infty - \mu_u^*\).
	
	We evaluated the bias across an extensive set of parameter combinations:
	\begin{itemize}
		\item Nearest neighbor order: \(\ell \in \{1, 2, 3, 4, 5, 6\}\),
		\item Number of angular sectors: \(q \in \{1, 2, 3, 4\}\),
		\item Population density: \(\lambda \in \{0.001, 0.006, 0.011, \dots, 2.001\}\) (step size 0.005),
		\item Aggregation parameter: \(k \in \{0.5, 1.0, 1.5, \dots, 10\}\) (step size 0.5),
		\item Truncation radius: \(C \in \{5, 10, 20, 30, 40, 50\}\),
		\item Moment order: \(u \in \{-2, -1, 1, 2\}\).
	\end{itemize}
	
	For all valid parameter combinations (satisfying \(k > u/2\)), the absolute asymptotic bias of \(\hat{\bE[R^u]}\) is strictly smaller than that of \(M_u\) (i.e., \(|B(\hat{\bE[R^u]})| < |B(M_u)|\)).

	\section{Performance of Poisson-based Estimators on Aggregated (Thomas Process) Populations}
	\label{sec:supp_poisson_on_thomas}
	
	This section provides the full simulation results for the four Poisson-based censored estimators, $\hat{\lambda}_{\text{DK}}$, $\hat{\lambda}_{C}^{(c)}$, $\hat{\lambda}_{P}^{(c)}$, and $\hat{\lambda}_{\text{MLE}}^{(c)}$, when applied to aggregated populations generated from a Thomas cluster process. These results complement the main-text comparison by showing how estimators derived under the complete spatial randomness assumption behave under increasing spatial aggregation.
	
\begin{figure}[!t]
		\centering
		\includegraphics[width=1\linewidth]{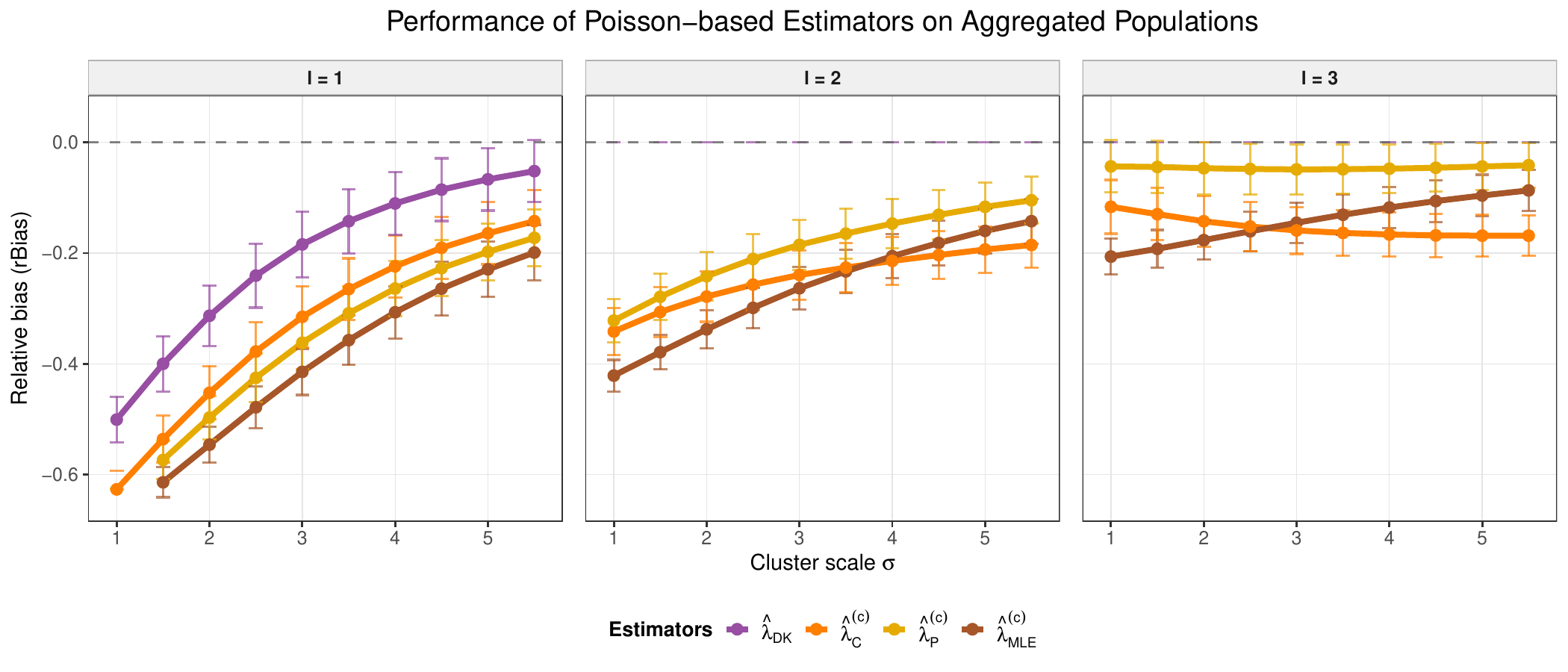} 
		\caption{
			Relative bias (points and solid lines) of the four Poisson-based censored estimators across a gradient of cluster scales ($\sigma$) for nearest-neighbour orders $\ell=1,2,$ and $3$. Error bars represent $\pm$ one relative standard deviation. Simulations are based on a Thomas cluster process with intensity $\lambda=0.05$ and censoring radius $C=10$ m.
		}
		\label{fig:poisson_estimators_supp}
	\end{figure}

	Figure~\ref{fig:poisson_estimators_supp} shows the relative bias of the four Poisson-based estimators across a gradient of cluster scales $\sigma$ for nearest-neighbour orders $\ell=1,2,$ and $3$. Across all scenarios, the four estimators exhibited consistent negative bias, and this bias became more pronounced as aggregation strengthened (i.e., as $\sigma$ decreased). These results confirm that Poisson-based estimators are systematically sensitive to departures from complete spatial randomness and therefore perform poorly when applied to clustered populations.
	
	\section{Empirical Performance Across Varying Censoring Radii and Forest Plots}
	\label{sec:supp_empirical}
	
	This section provides a comprehensive summary of the performance of all seven censored density estimators across the full range of maximum search radii ($C = 10, 20, 30, 40$ m) in both the Barro Colorado Island (BCI) 50‑ha plot and the Harvard Forest 35‑ha plot. The main text presents, for the BCI plot with $C = 20$ m, the differences in absolute relative bias and relative RMSE relative to the NBD‑based MLE. Here we supplement those results with three complementary views: (i) the raw $\text{rBias}$ and $\text{rRMSE}$ boxplots for all $C$ and $\ell$, (ii) the absolute differences in $|\text{rBias}|$ and $\text{rRMSE}$ relative to $\hat{\lambda}_{\text{n,MLE}}^{(c)}$, and (iii) the relative improvements, calculated as $(|\cdot|_{\text{method}} - |\cdot|_{\text{NBD-MLE}}) / |\cdot|_{\text{method}}$. All three types of plots are shown for both the BCI and Harvard Forest plots.
	
	
	\subsection{BCI plot: raw performance metrics.}
\begin{figure}[!t]
		\centering
		\includegraphics[width=1\linewidth]{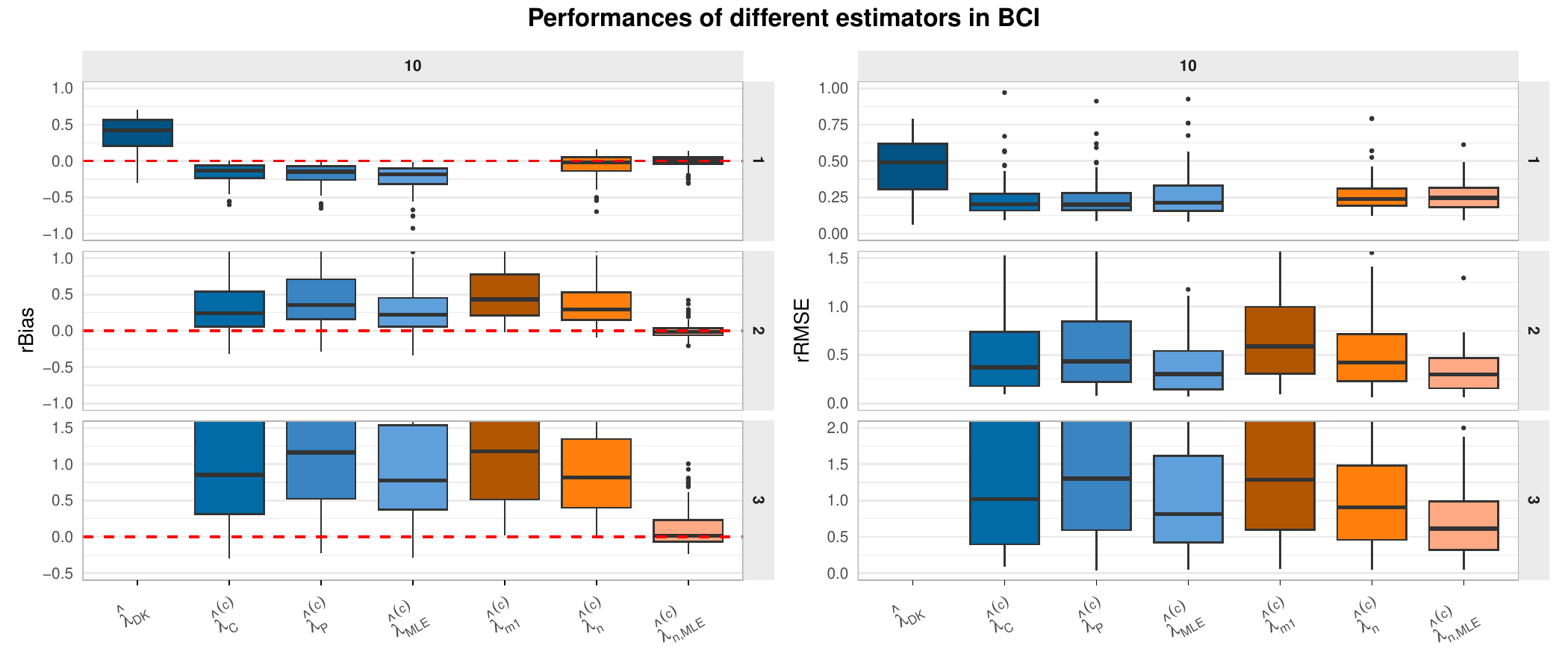}
		\caption{Performance of the seven estimators for the BCI plot with $C = 10$\,m. Left column: $\text{rBias}$; Right column: $\text{rRMSE}$. Rows from top to bottom: $\ell = 1, 2, 3$.}
		\label{fig:bci_c10}
	\end{figure}

\begin{figure}[!t]
		\centering
		\includegraphics[width=1\linewidth]{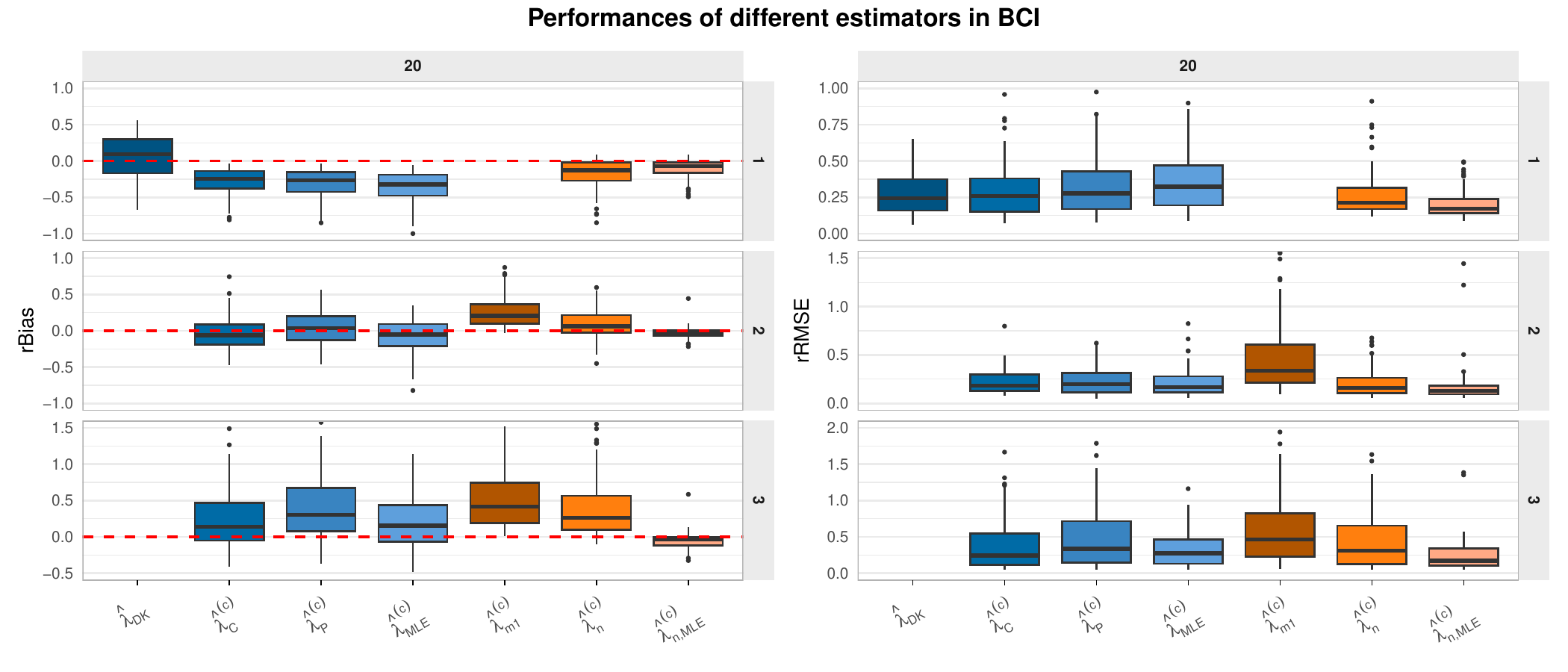}
		\caption{Performance of the seven estimators for the BCI plot with $C = 20$\,m. Layout identical to Figure~\ref{fig:bci_c10}.}
		\label{fig:bci_c20}
	\end{figure}

\begin{figure}[!t]
		\centering
		\includegraphics[width=1\linewidth]{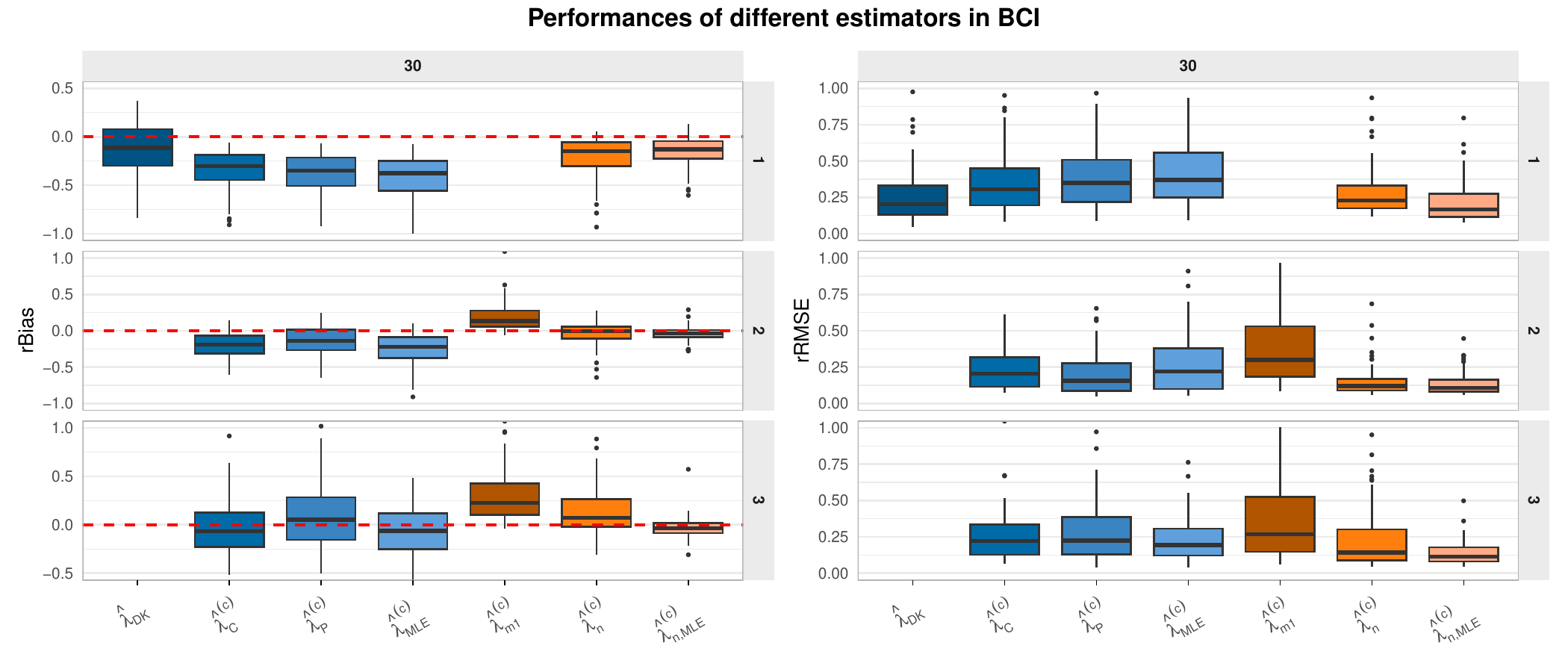}
		\caption{Performance of the seven estimators for the BCI plot with $C = 30$\,m. Layout identical to Figure~\ref{fig:bci_c10}.}
		\label{fig:bci_c30}
	\end{figure}

\begin{figure}[!t]
		\centering
		\includegraphics[width=1\linewidth]{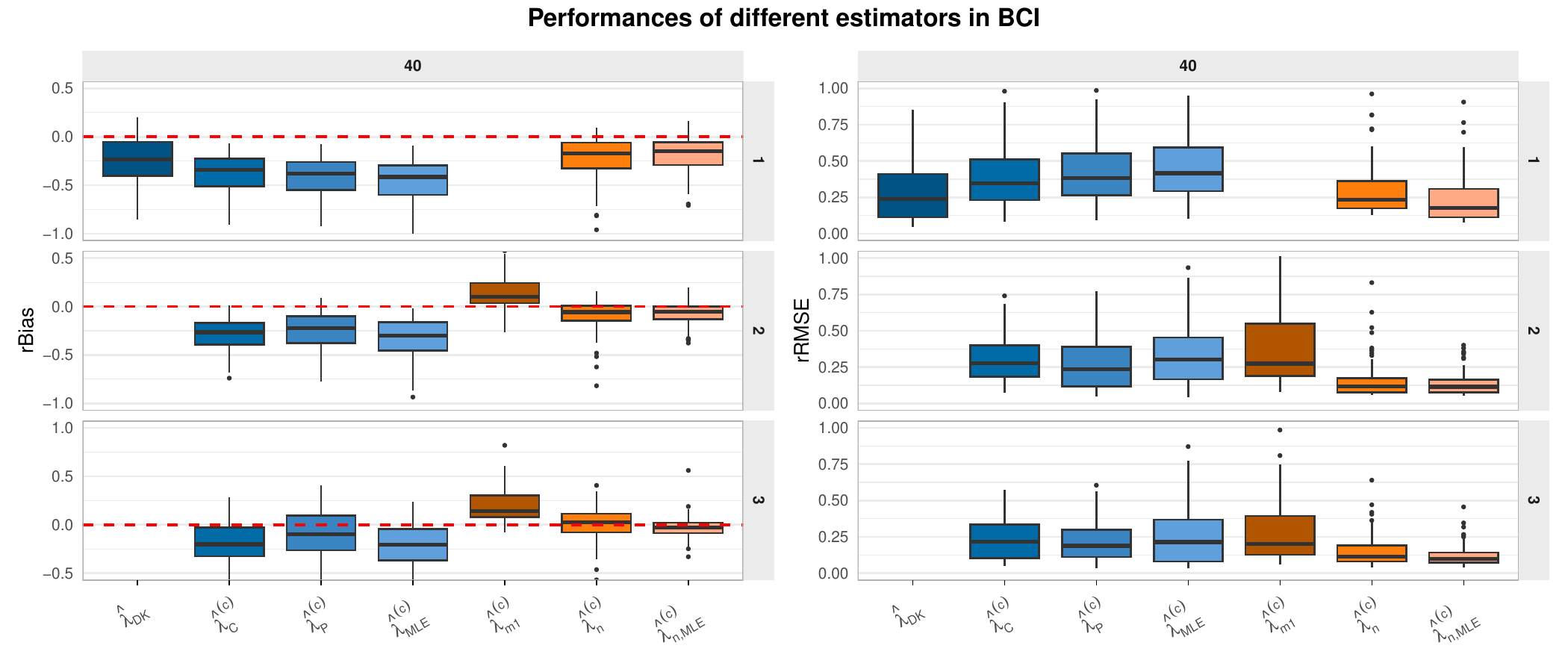}
		\caption{Performance of the seven estimators for the BCI plot with $C = 40$\,m. Layout identical to Figure~\ref{fig:bci_c10}.}
		\label{fig:bci_c40}
	\end{figure}

	Figures~\ref{fig:bci_c10}--\ref{fig:bci_c40} display the distribution of $\text{rBias}$ and $\text{rRMSE}$ for the seven estimators at $C = 10, 20, 30$ and $40$ m. In all cases the estimator performance varies systematically with nearest‑neighbour order $\ell$, and the NBD‑based estimators (particularly $\hat{\lambda}_{\text{n,MLE}}^{(c)}$ and $\hat{\lambda}_{n}^{(c)}$) consistently achieve the lowest median errors.
	
	\subsection{BCI plot: absolute differences relative to $\hat{\lambda}_{\text{n,MLE}}^{(c)}$.}
\begin{figure}[!t]
		\centering
		\includegraphics[width=1\linewidth]{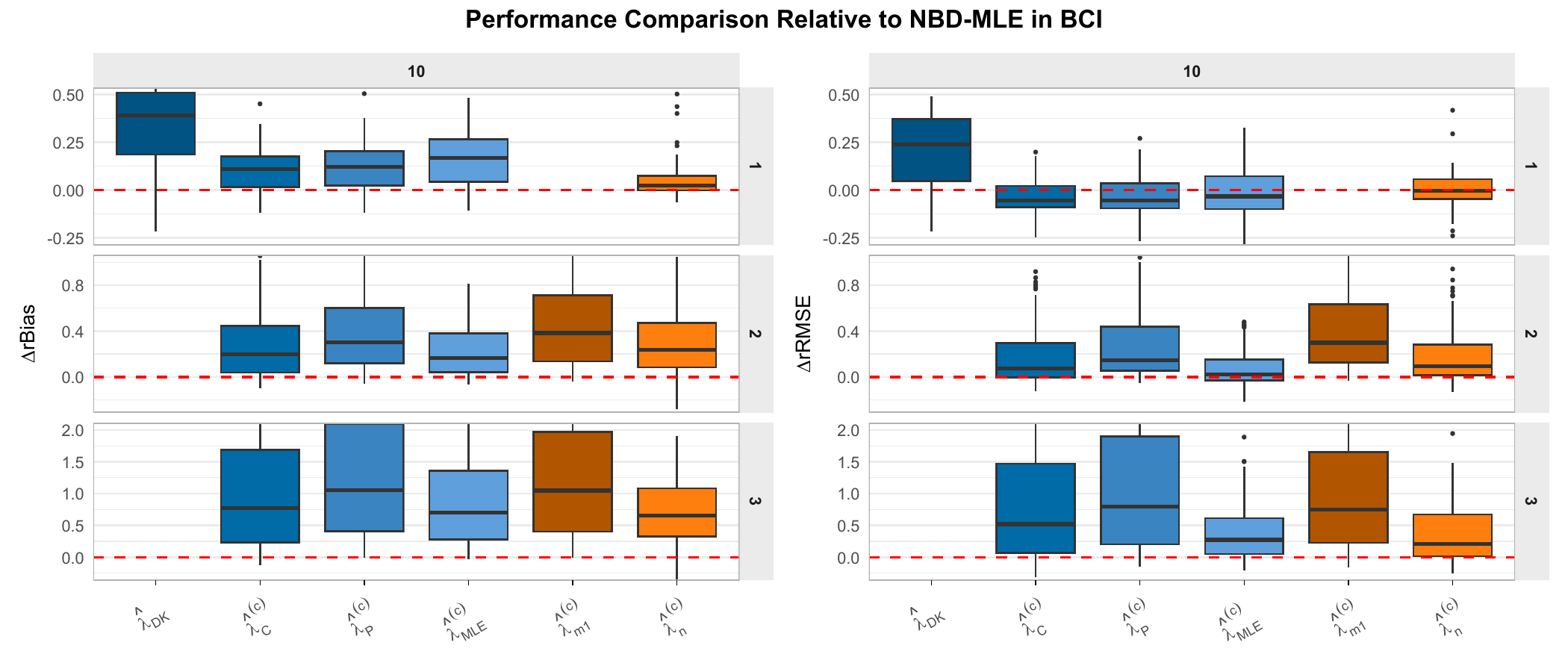}
		\caption{Differences in $|\text{rBias}|$ (left) and $\text{rRMSE}$ (right) between each method and $\hat{\lambda}_{\text{n,MLE}}^{(c)}$ for the BCI plot with $C = 10$\,m and $\ell = 1,2,3$. Layout as in the main‑text Figure~2.}
		\label{fig:new_bci_c10}
	\end{figure}

\begin{figure}[!t]
		\centering
		\includegraphics[width=1\linewidth]{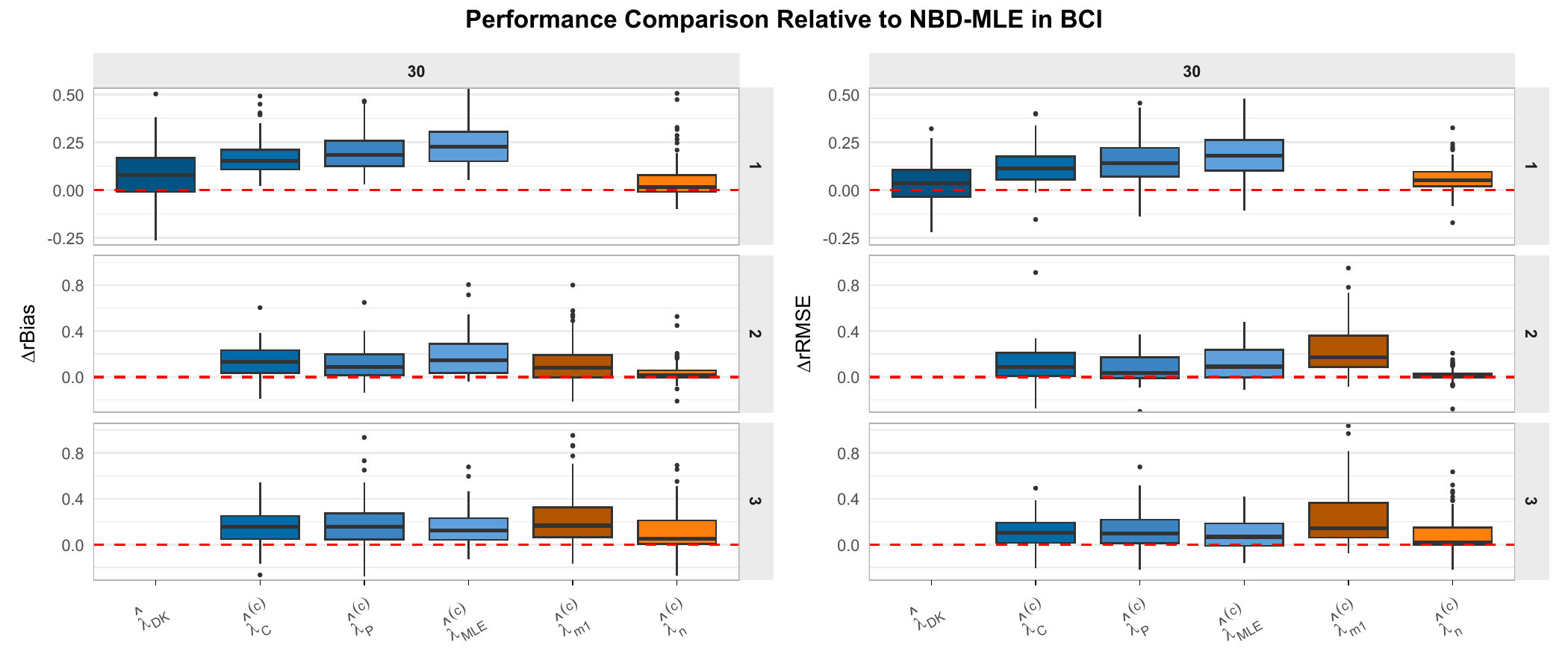}
		\caption{Differences in $|\text{rBias}|$ (left) and $\text{rRMSE}$ (right) between each method and $\hat{\lambda}_{\text{n,MLE}}^{(c)}$ for the BCI plot with $C = 30$\,m and $\ell = 1,2,3$. Layout as in the main‑text Figure~2.}
		\label{fig:new_bci_c30}
	\end{figure}

\begin{figure}[!t]
		\centering
		\includegraphics[width=1\linewidth]{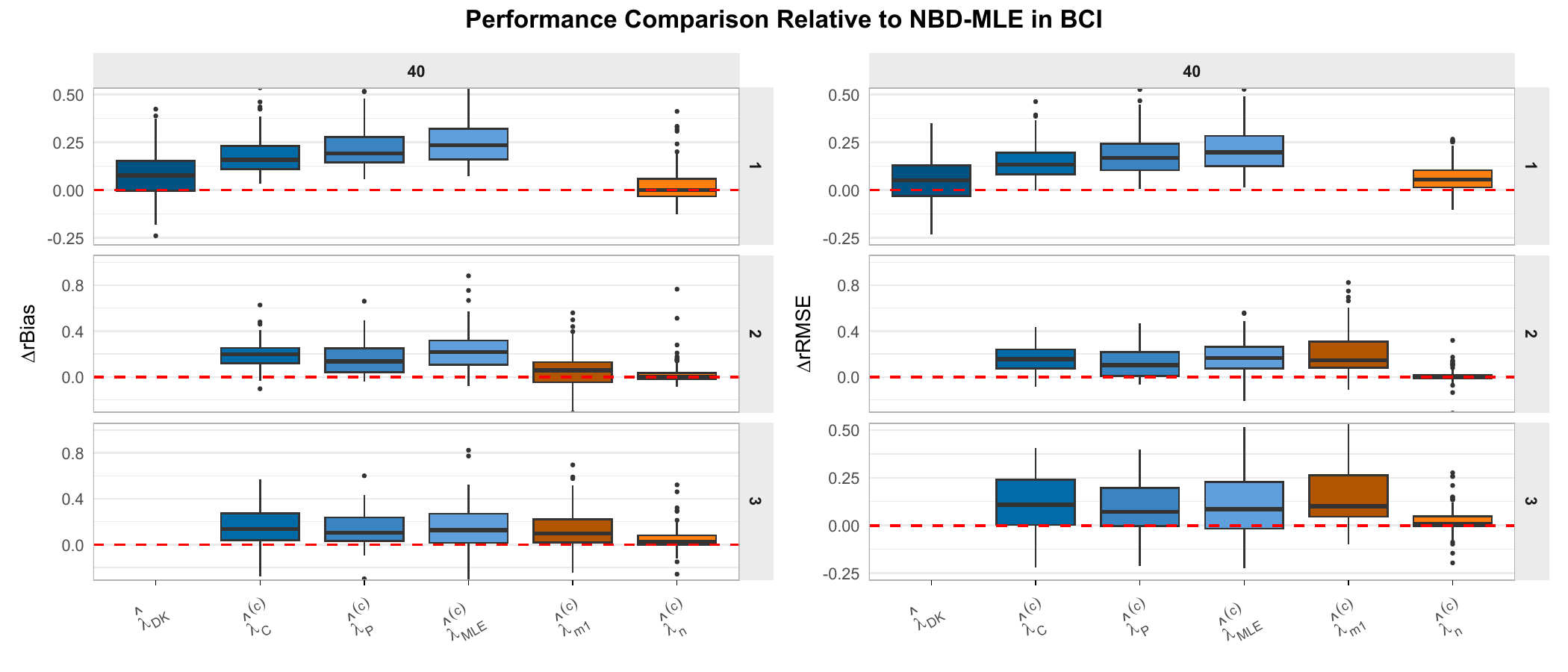}
		\caption{Differences in $|\text{rBias}|$ (left) and $\text{rRMSE}$ (right) between each method and $\hat{\lambda}_{\text{n,MLE}}^{(c)}$ for the BCI plot with $C = 40$\,m and $\ell = 1,2,3$. Layout as in the main‑text Figure~2.}
		\label{fig:new_bci_c40}
	\end{figure}

	To directly quantify the advantage of $\hat{\lambda}_{\text{n,MLE}}^{(c)}$ over the other six estimators, Figures~\ref{fig:new_bci_c10}--\ref{fig:new_bci_c40} show the differences in $|\text{rBias}|$ and $\text{rRMSE}$ relative to the NBD-MLE for $C = 10, 30,$ and $40$ m (the case $C = 20$ m appears in the main text). Across all radii, the majority of boxplots lie above zero, confirming that $\hat{\lambda}_{\text{n,MLE}}^{(c)}$ yields the smallest absolute errors for almost all species and values of $\ell$.
	
	\subsection{BCI plot: relative improvements over $\hat{\lambda}_{\text{n,MLE}}^{(c)}$.}
\begin{figure}[!t]
		\centering
		\includegraphics[width=1\linewidth]{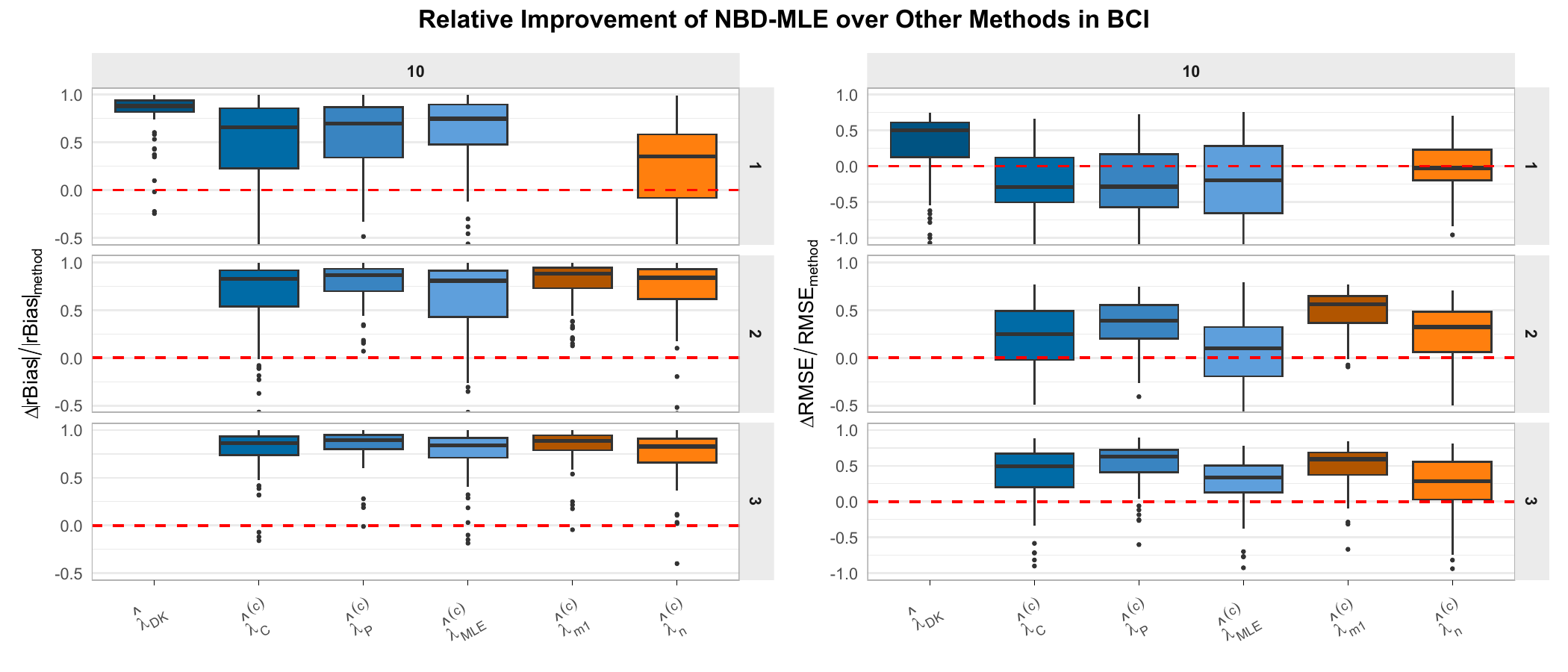}
		\caption{Relative improvement in $|\text{rBias}|$ (left) and $\text{rRMSE}$ (right) for each method compared to $\hat{\lambda}_{\text{n,MLE}}^{(c)}$ for the BCI plot with $C = 10$\,m and $\ell = 1,2,3$. Values above zero indicate that NBD-MLE outperforms the comparator.}
		\label{fig:latest_bci_c10}
	\end{figure}

\begin{figure}[!t]
		\centering
		\includegraphics[width=1\linewidth]{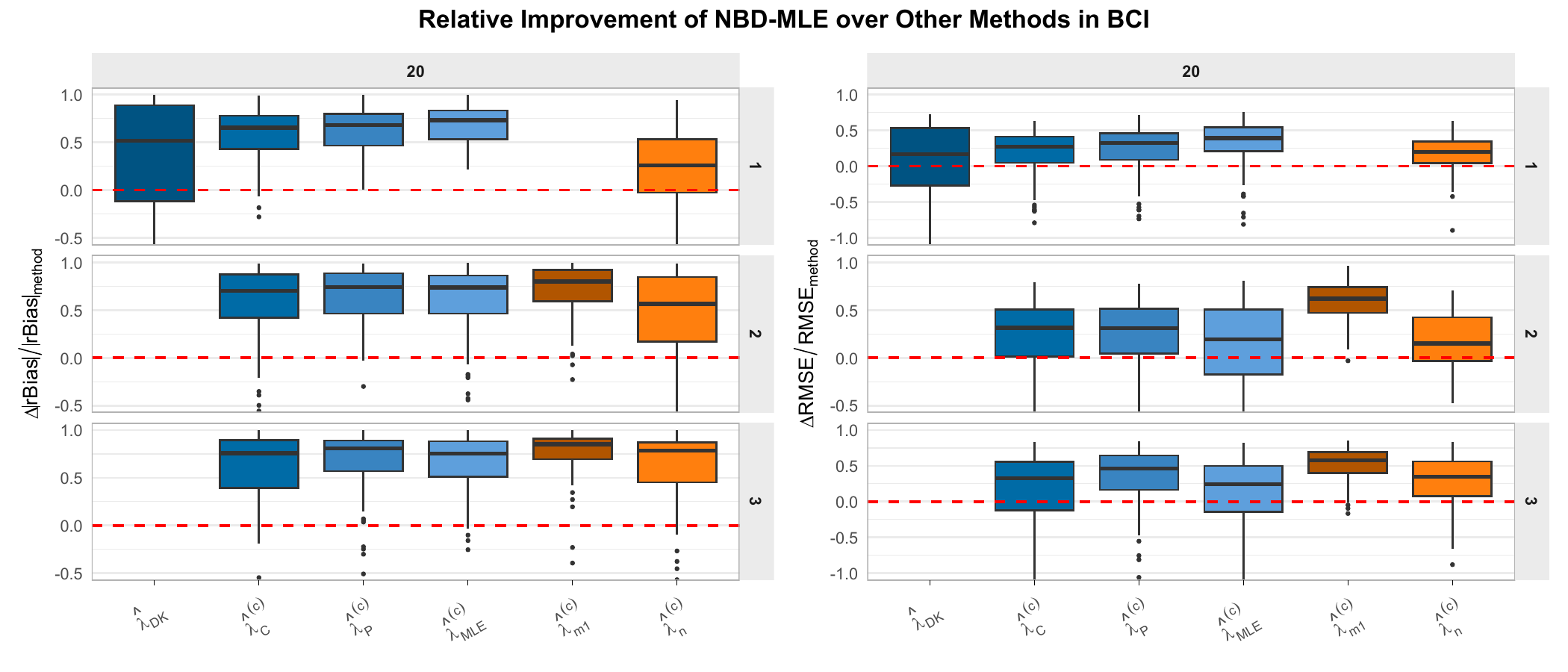}
		\caption{Same as Figure~\ref{fig:latest_bci_c10} but for $C = 20$\,m.}
		\label{fig:latest_bci_c20}
	\end{figure}

\begin{figure}[!t]
		\centering
		\includegraphics[width=1\linewidth]{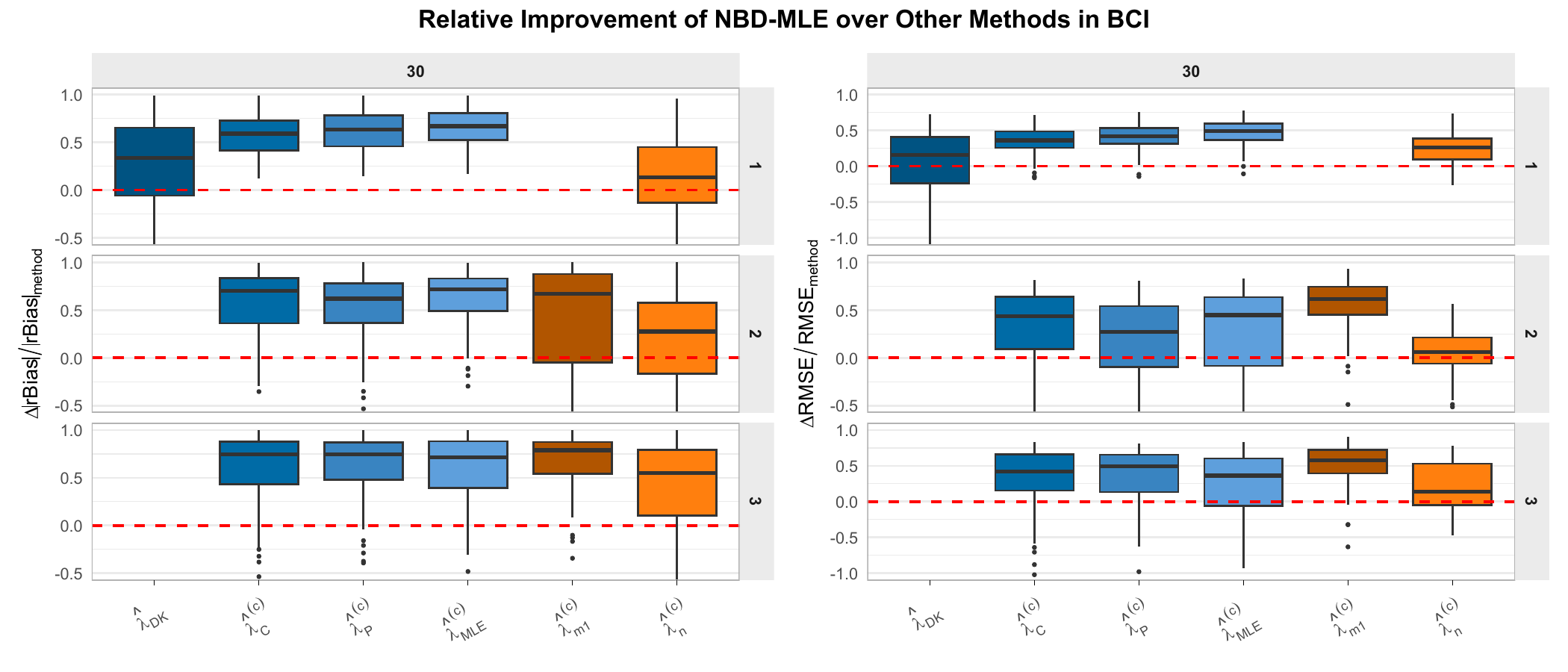}
		\caption{Same as Figure~\ref{fig:latest_bci_c10} but for $C = 30$\,m.}
		\label{fig:latest_bci_c30}
	\end{figure}

\begin{figure}[!t]
		\centering
		\includegraphics[width=1\linewidth]{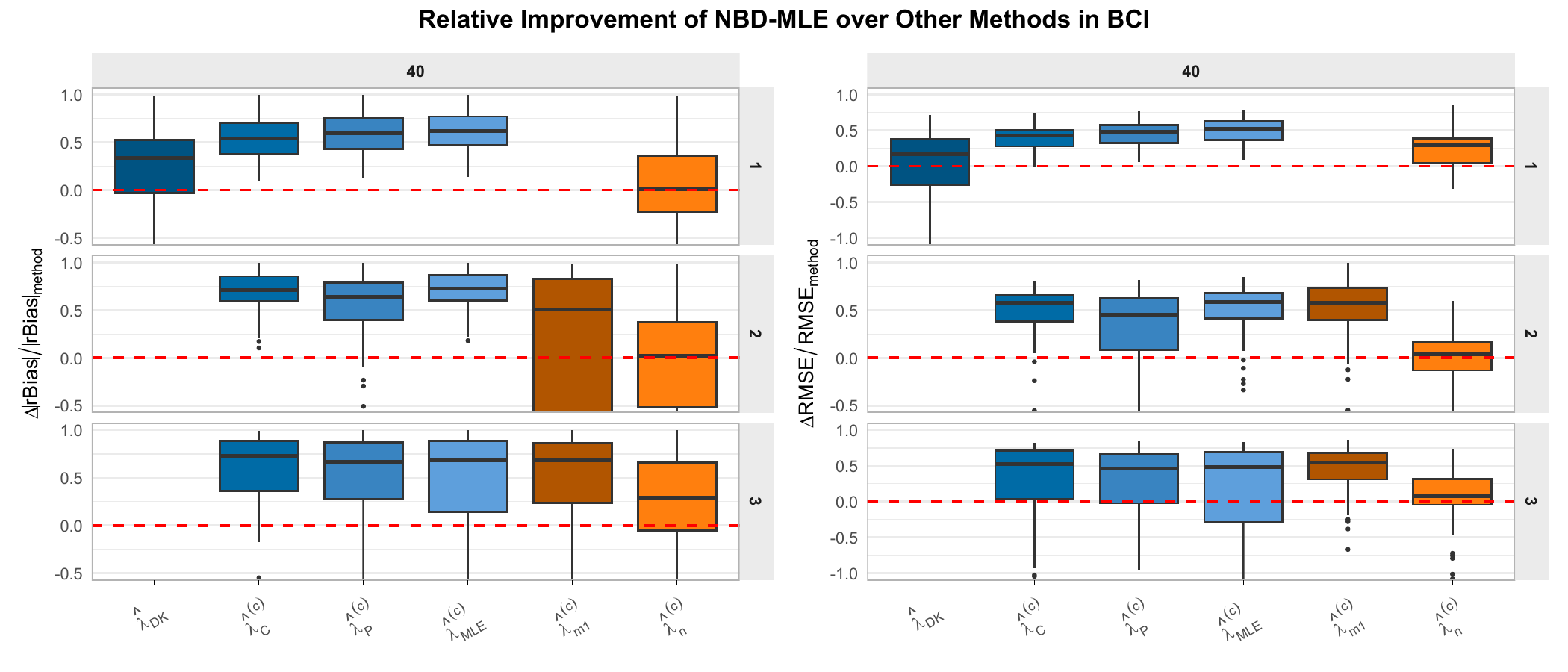}
		\caption{Same as Figure~\ref{fig:latest_bci_c10} but for $C = 40$\,m.}
		\label{fig:latest_bci_c40}
	\end{figure}

	Figures~\ref{fig:latest_bci_c10}--\ref{fig:latest_bci_c40} display the relative improvement of each method with respect to $\hat{\lambda}_{\text{n,MLE}}^{(c)}$, defined as $(|\text{rBias}|_{\text{method}} - |\text{rBias}|_{\text{NBD-MLE}}) / |\text{rBias}|_{\text{method}}$ (and analogously for $\text{rRMSE}$). Positive values indicate that the NBD-MLE reduces the error relative to the alternative method; the red dashed line at zero marks equal performance. The patterns reinforce the findings from the absolute differences: $\hat{\lambda}_{\text{n,MLE}}^{(c)}$ achieves a systematic and often substantial relative gain across species .
	
	\subsection{Harvard Forest plot: raw performance metrics.}
\begin{figure}[!t]
		\centering
		\includegraphics[width=1\linewidth]{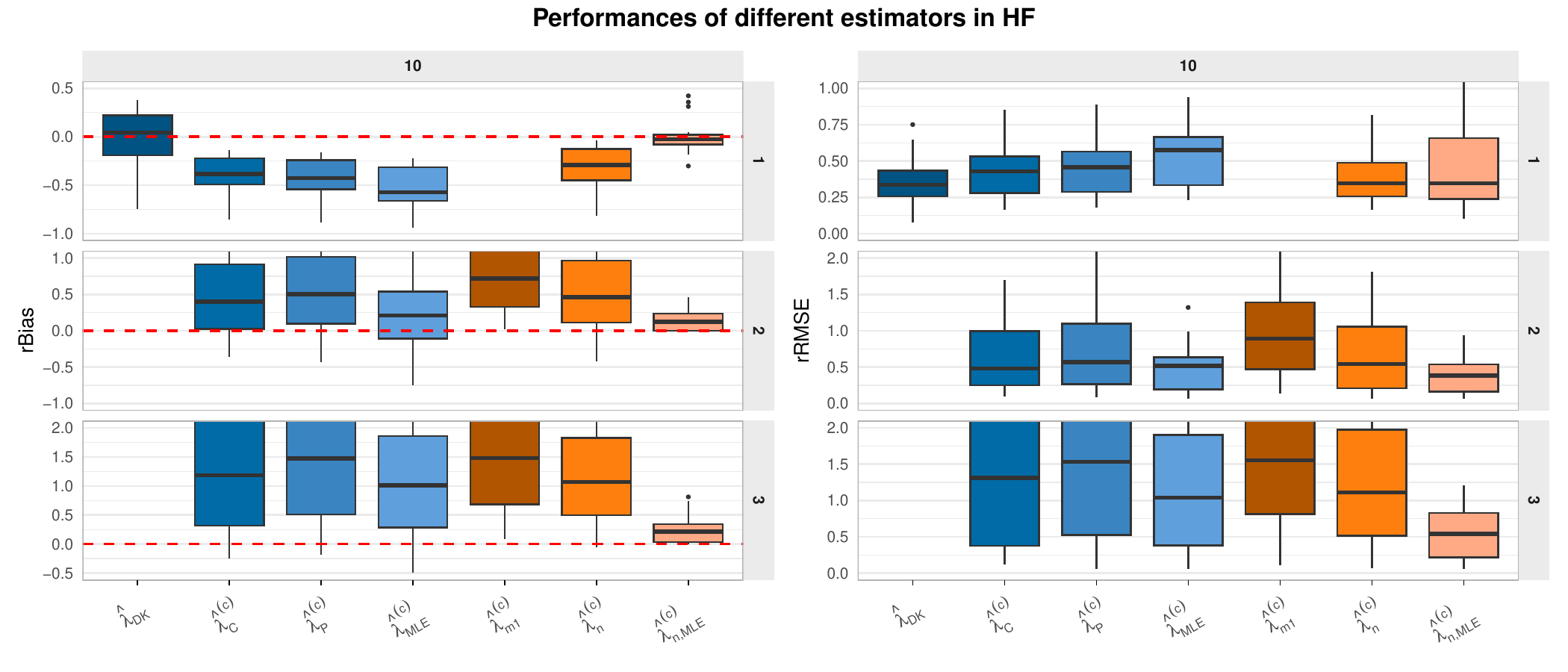}
		\caption{Performance of the seven estimators for the Harvard Forest plot with $C = 10$\,m. Layout identical to Figure~\ref{fig:bci_c10}.}
		\label{fig:harvard_c10}
	\end{figure}

\begin{figure}[!t]
		\centering
		\includegraphics[width=1\linewidth]{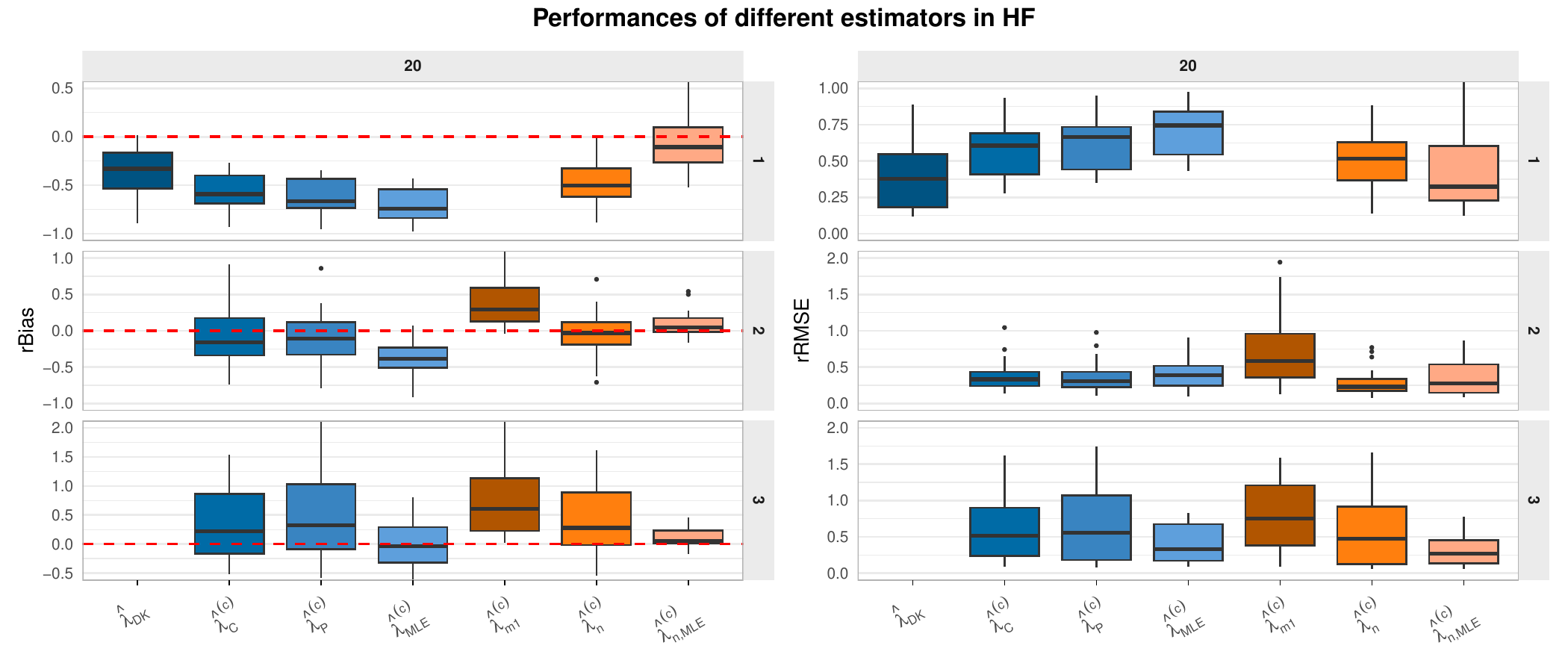}
		\caption{Performance of the seven estimators for the Harvard Forest plot with $C = 20$\,m. Layout identical to Figure~\ref{fig:bci_c10}.}
		\label{fig:harvard_c20}
	\end{figure}

\begin{figure}[!t]
		\centering
		\includegraphics[width=1\linewidth]{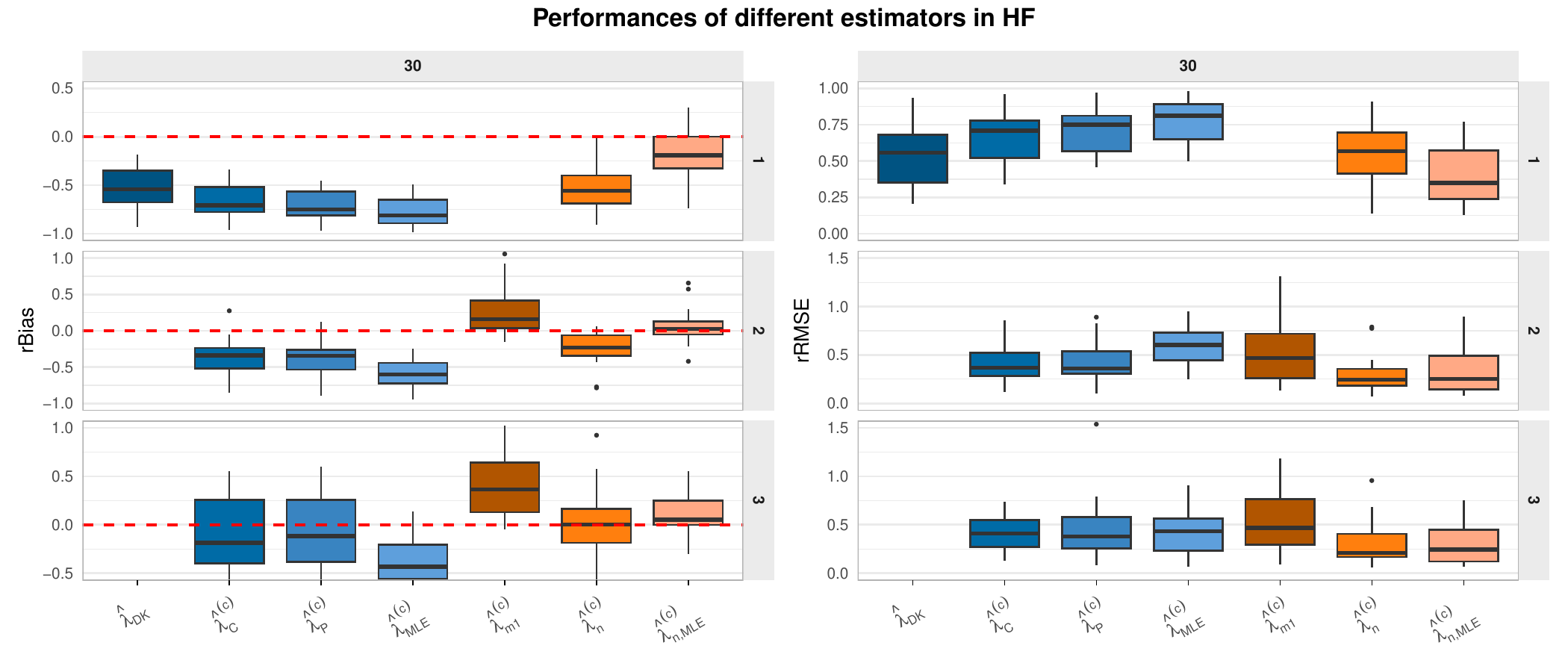}
		\caption{Performance of the seven estimators for the Harvard Forest plot with $C = 30$\,m. Layout identical to Figure~\ref{fig:bci_c10}.}
		\label{fig:harvard_c30}
	\end{figure}

\begin{figure}[!t]
		\centering
		\includegraphics[width=1\linewidth]{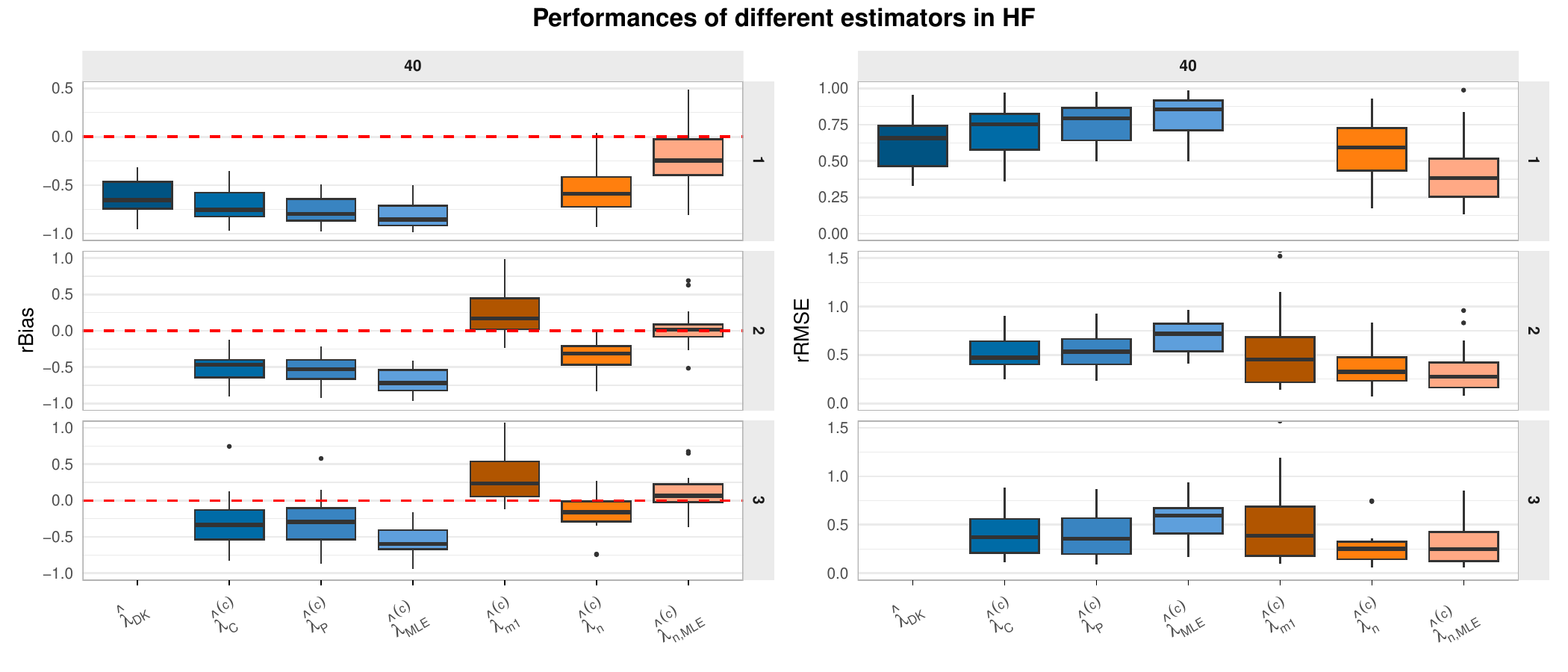}
		\caption{Performance of the seven estimators for the Harvard Forest plot with $C = 40$\,m. Layout identical to Figure~\ref{fig:bci_c10}.}
		\label{fig:harvard_c40}
	\end{figure}

	Figures~\ref{fig:harvard_c10}--\ref{fig:harvard_c40} present the raw $\text{rBias}$ and $\text{rRMSE}$ for the Harvard Forest plot across all four censoring radii.
	
	\subsection{Harvard Forest plot: absolute differences relative to $\hat{\lambda}_{\text{n,MLE}}^{(c)}$.}
\begin{figure}[!t]
		\centering
		\includegraphics[width=1\linewidth]{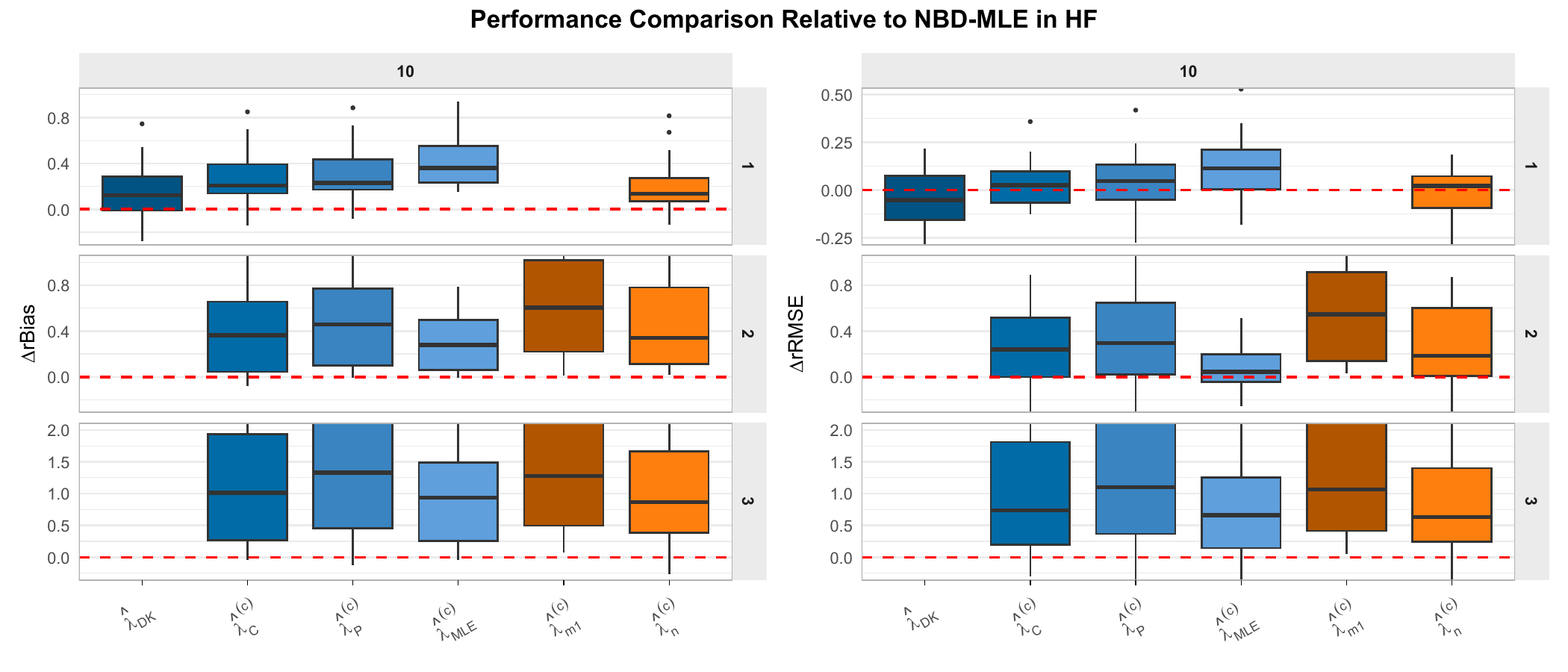}
		\caption{Differences in $|\text{rBias}|$ (left) and $\text{rRMSE}$ (right) between each method and $\hat{\lambda}_{\text{n,MLE}}^{(c)}$ for the Harvard Forest plot with $C = 10$\,m and $\ell = 1,2,3$. Layout as in the main‑text Figure~2.}
		\label{fig:new_harvard_c10}
	\end{figure}

\begin{figure}[!t]
		\centering
		\includegraphics[width=1\linewidth]{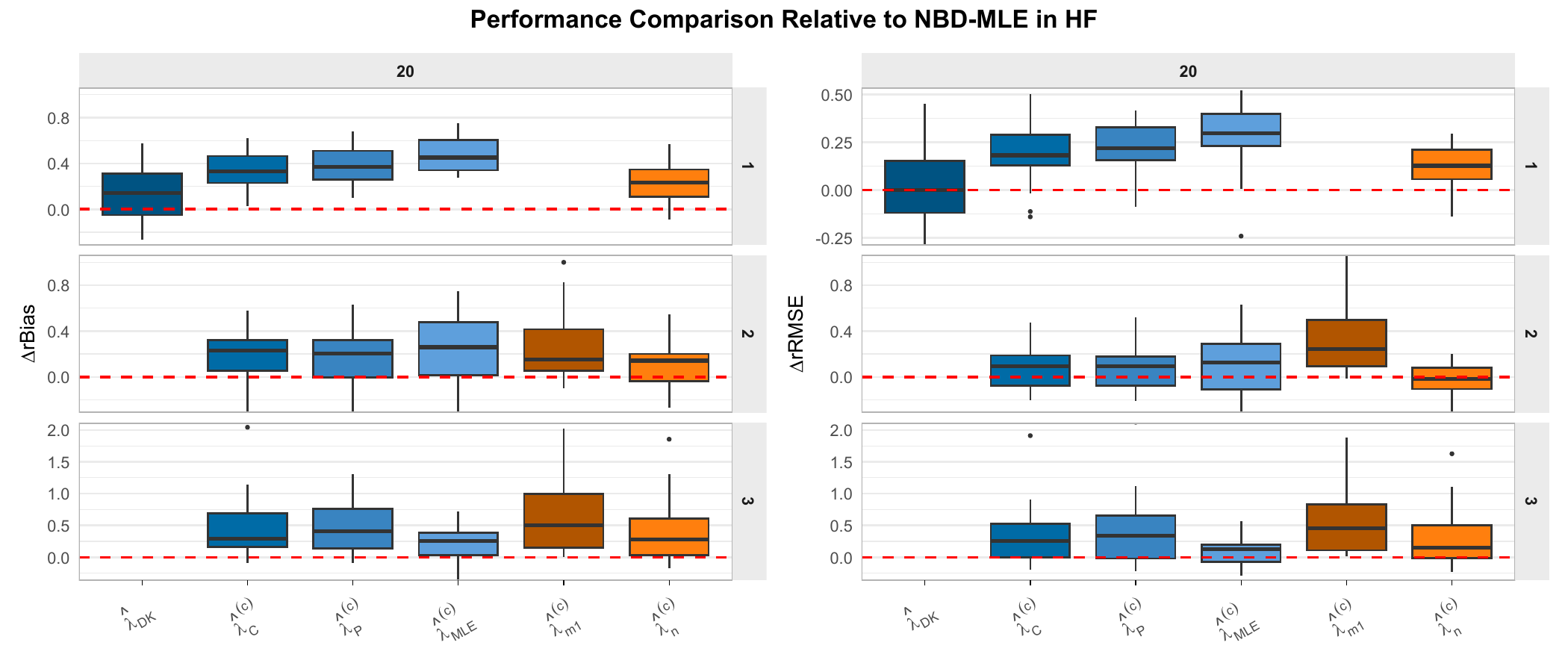}
		\caption{Differences in $|\text{rBias}|$ (left) and $\text{rRMSE}$ (right) between each method and $\hat{\lambda}_{\text{n,MLE}}^{(c)}$ for the Harvard Forest plot with $C = 20$\,m and $\ell = 1,2,3$. Layout as in the main‑text Figure~2.}
		\label{fig:new_harvard_c20}
	\end{figure}

\begin{figure}[!t]
		\centering
		\includegraphics[width=1\linewidth]{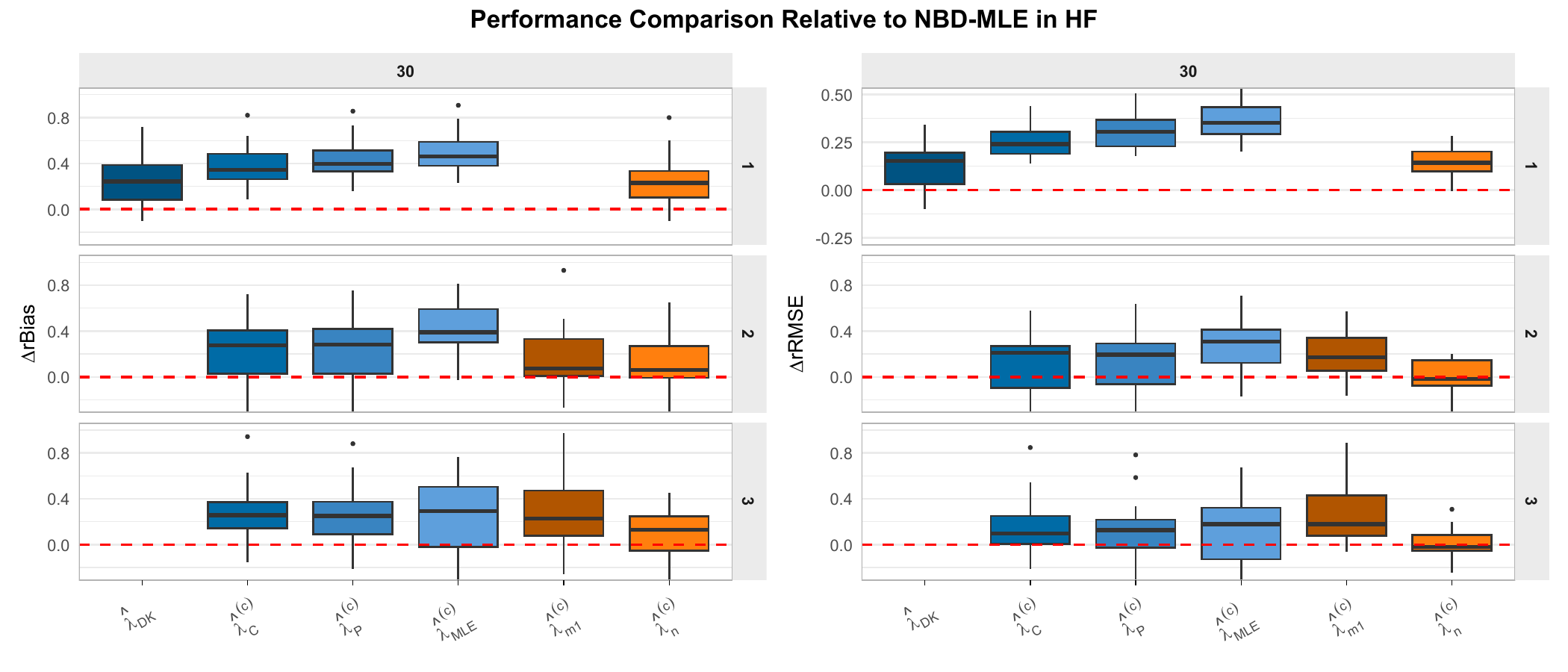}
		\caption{Differences in $|\text{rBias}|$ (left) and $\text{rRMSE}$ (right) between each method and $\hat{\lambda}_{\text{n,MLE}}^{(c)}$ for the Harvard Forest plot with $C = 30$\,m and $\ell = 1,2,3$. Layout as in the main‑text Figure~2.}
		\label{fig:new_harvard_c30}
	\end{figure}

\begin{figure}[!t]
		\centering
		\includegraphics[width=1\linewidth]{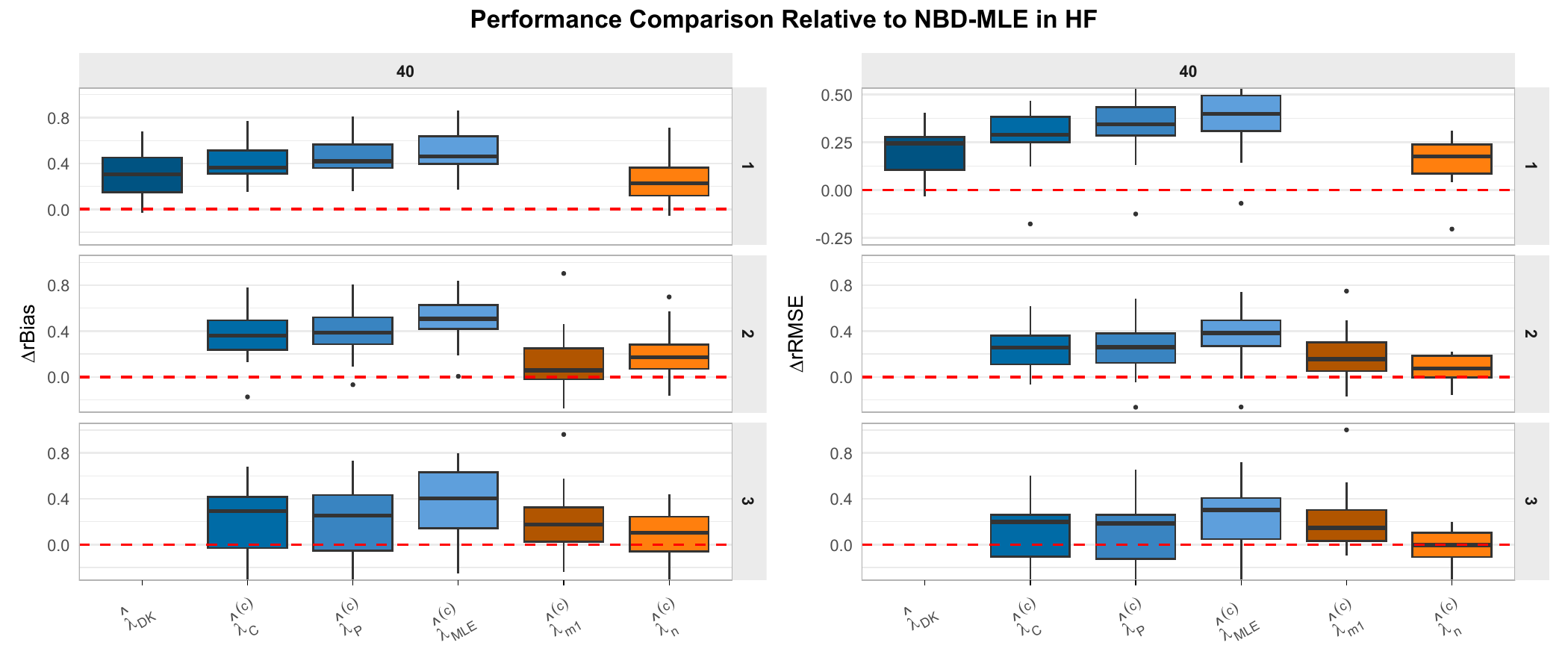}
		\caption{Differences in $|\text{rBias}|$ (left) and $\text{rRMSE}$ (right) between each method and $\hat{\lambda}_{\text{n,MLE}}^{(c)}$ for the Harvard Forest plot with $C = 40$\,m and $\ell = 1,2,3$.}
		\label{fig:new_harvard_c40}
	\end{figure}

	Figures~\ref{fig:new_harvard_c10}--\ref{fig:new_harvard_c40} display the corresponding absolute differences from $\hat{\lambda}_{\text{n,MLE}}^{(c)}$. The patterns are broadly consistent with those observed at BCI: $\hat{\lambda}_{\text{n,MLE}}^{(c)}$ remains the most accurate estimator overall.
	
	\subsection{Harvard Forest plot: relative improvements over $\hat{\lambda}_{\text{n,MLE}}^{(c)}$.}
\begin{figure}[!t]
		\centering
		\includegraphics[width=1\linewidth]{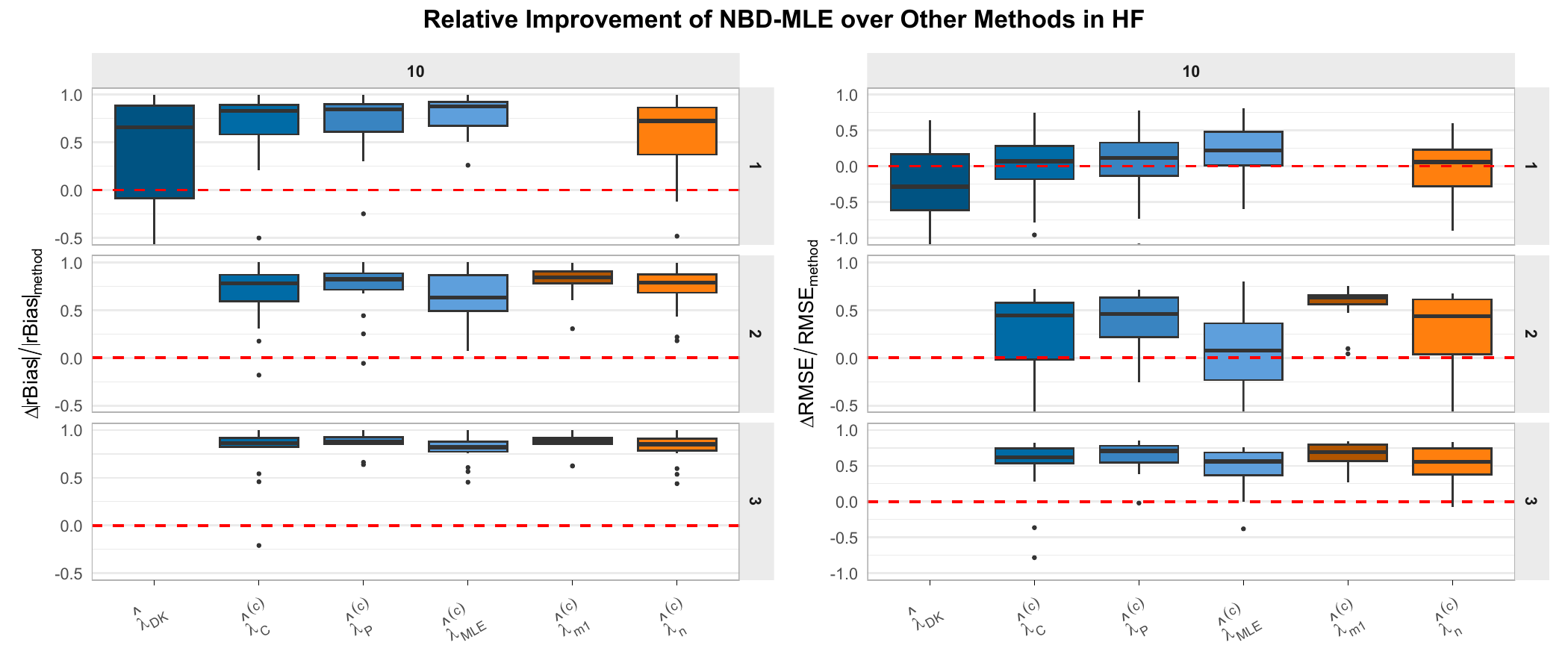}
		\caption{Relative improvement in $|\text{rBias}|$ (left) and $\text{rRMSE}$ (right) for each method compared to $\hat{\lambda}_{\text{n,MLE}}^{(c)}$ for the Harvard Forest plot with $C = 10$\,m and $\ell = 1,2,3$.}
		\label{fig:latest_harvard_c10}
	\end{figure}

\begin{figure}[!t]
		\centering
		\includegraphics[width=1\linewidth]{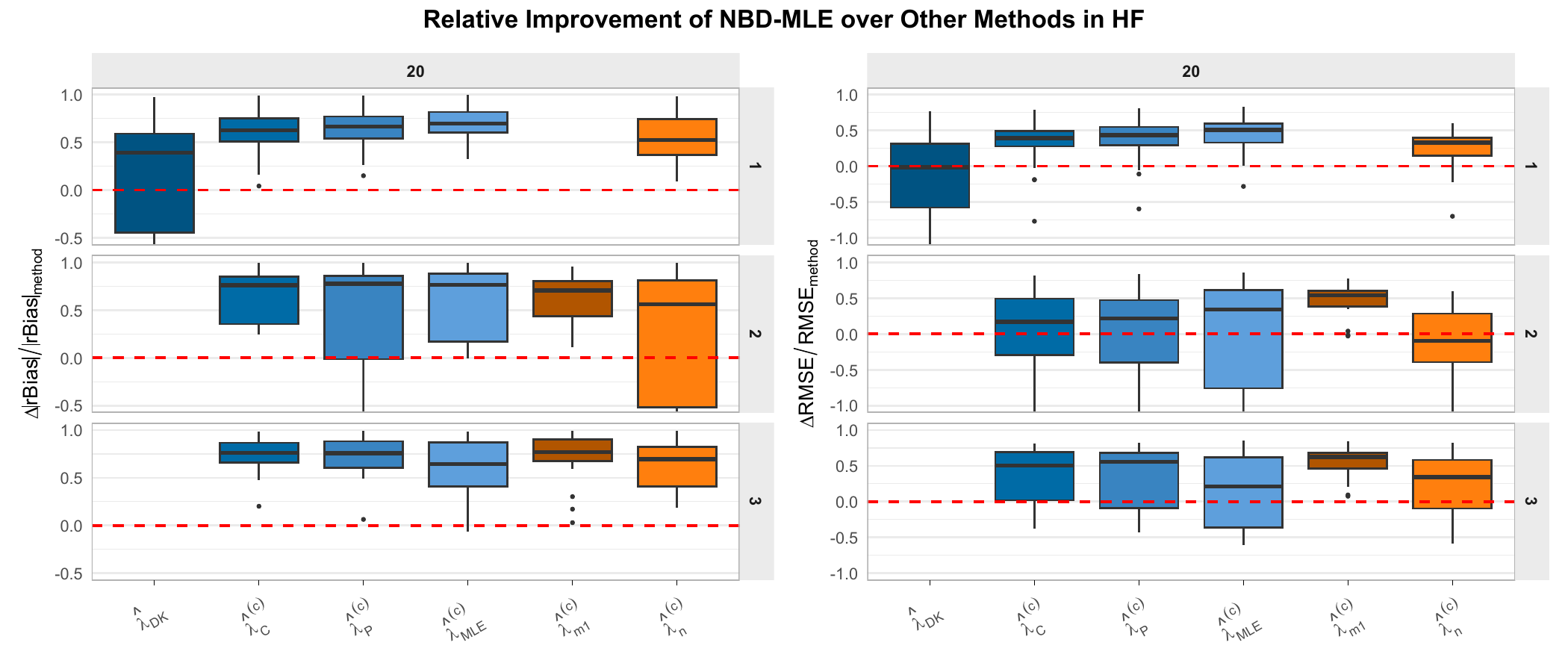}
		\caption{Same as Figure~\ref{fig:latest_harvard_c10} but for $C = 20$\,m.}
		\label{fig:latest_harvard_c20}
	\end{figure}

\begin{figure}[!t]
		\centering
		\includegraphics[width=1\linewidth]{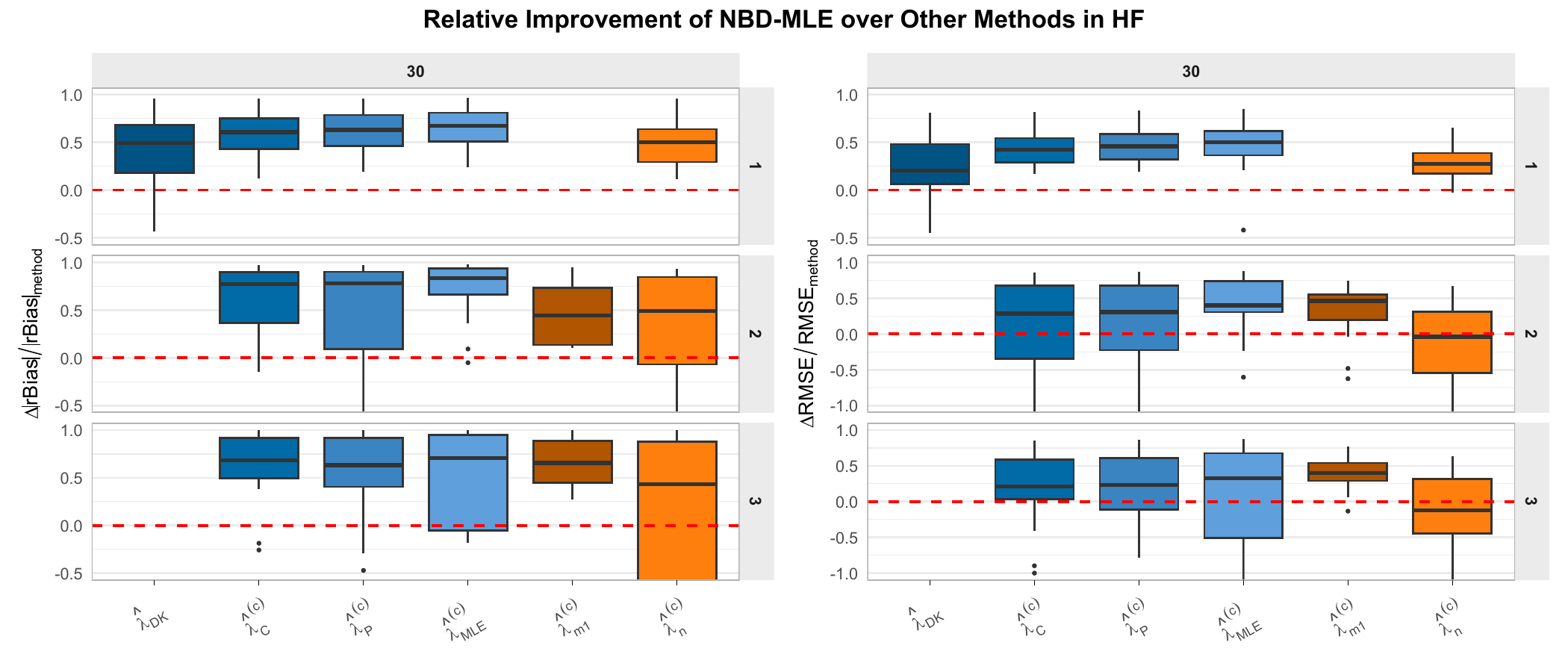}
		\caption{Same as Figure~\ref{fig:latest_harvard_c10} but for $C = 30$\,m.}
		\label{fig:latest_harvard_c30}
	\end{figure}

\begin{figure}[!t]
		\centering
		\includegraphics[width=1\linewidth]{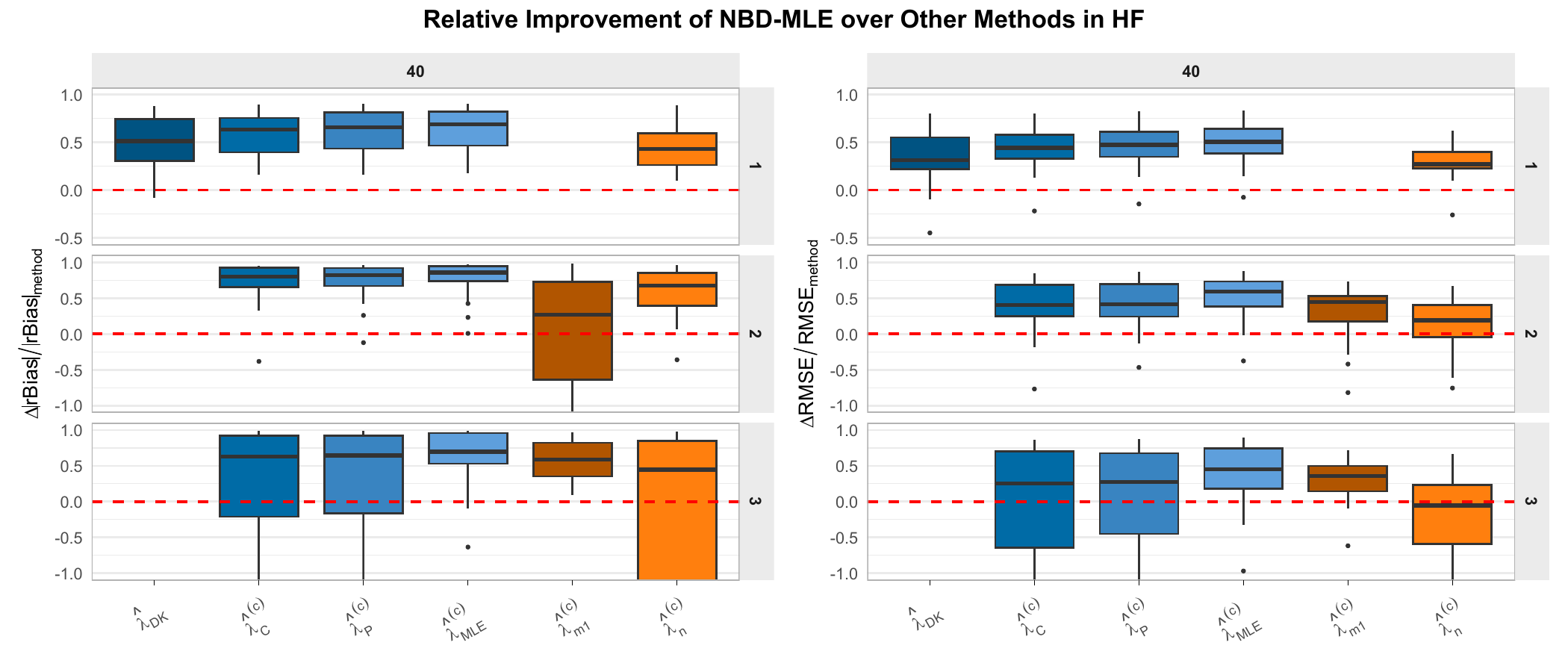}
		\caption{Same as Figure~\ref{fig:latest_harvard_c10} but for $C = 40$\,m.}
		\label{fig:latest_harvard_c40}
	\end{figure}

	Figures~\ref{fig:latest_harvard_c10}--\ref{fig:latest_harvard_c40} show the relative improvements for the Harvard Forest data. As with BCI, the NBD-MLE yields consistent positive relative gains, confirming its robust performance.
	
	\subsection{Summary of absolute and relative improvement of \texorpdfstring{$\hat{\lambda}_{\text{n,MLE}}^{(c)}$}{NBD-MLE} over existing estimators (\texorpdfstring{$\ell=1$}{l=1})}
	
	Tables~\ref{tab:abs_improve} and \ref{tab:rel_improve} summarise the distribution
	(Q1, median, Q3) of the absolute and relative improvement in $|\text{rBias}|$ achieved by
	the NBD-based maximum likelihood estimator $\hat{\lambda}_{\text{n,MLE}}^{(c)}$ over the
	two classical corrections designed for $\ell=1$, namely $\hat{\lambda}_{\text{DK}}$ and
	$\hat{\lambda}_{C}^{(c)}$. 
	Q1 and Q3 denote the 25th and 75th percentiles of the across‑species distributions of
	these improvements.
	Positive values indicate that $\hat{\lambda}_{\text{n,MLE}}^{(c)}$ yields a smaller
	absolute relative bias than the comparator.
	
	\begin{table}[!ht]
		\centering
		\caption{Absolute improvement in $|\text{rBias}|$: $\Delta|\text{rBias}| = |\text{rBias}|_{\text{other}} - |\text{rBias}|_{\text{NBD-MLE}}$.}
		\label{tab:abs_improve}
		\small
		\begin{tabular}{l c l c c c}
			\toprule
			Plot & $C$ (m) & Estimator & Q1 & Median & Q3 \\
			\midrule
			\multirow{8}{*}{BCI} 
			& 10 & $\hat{\lambda}_{C}^{(c)}$      & 0.015 & 0.110 & 0.177 \\
			& 10 & $\hat{\lambda}_{\text{DK}}$    & 0.186 & 0.393 & 0.511 \\
			& 20 & $\hat{\lambda}_{C}^{(c)}$      & 0.095 & 0.140 & 0.199 \\
			& 20 & $\hat{\lambda}_{\text{DK}}$    & -0.018& 0.096 & 0.276 \\
			& 30 & $\hat{\lambda}_{C}^{(c)}$      & 0.108 & 0.154 & 0.213 \\
			& 30 & $\hat{\lambda}_{\text{DK}}$    & -0.010& 0.078 & 0.169 \\
			& 40 & $\hat{\lambda}_{C}^{(c)}$      & 0.109 & 0.159 & 0.231 \\
			& 40 & $\hat{\lambda}_{\text{DK}}$    & -0.004& 0.076 & 0.153 \\
			\cmidrule{1-6}
			\multirow{8}{*}{HF}  
			& 10 & $\hat{\lambda}_{C}^{(c)}$      & 0.139 & 0.208 & 0.393 \\
			& 10 & $\hat{\lambda}_{\text{DK}}$    & -0.012& 0.120 & 0.283 \\
			& 20 & $\hat{\lambda}_{C}^{(c)}$      & 0.230 & 0.331 & 0.465 \\
			& 20 & $\hat{\lambda}_{\text{DK}}$    & -0.050& 0.141 & 0.310 \\
			& 30 & $\hat{\lambda}_{C}^{(c)}$      & 0.263 & 0.344 & 0.483 \\
			& 30 & $\hat{\lambda}_{\text{DK}}$    & 0.085 & 0.243 & 0.386 \\
			& 40 & $\hat{\lambda}_{C}^{(c)}$      & 0.311 & 0.363 & 0.515 \\
			& 40 & $\hat{\lambda}_{\text{DK}}$    & 0.147 & 0.303 & 0.451 \\
			\bottomrule
		\end{tabular}
	\end{table}
	
	\begin{table}[!ht]
		\centering
		\caption{Relative improvement in $|\text{rBias}|$: $(|\text{rBias}|_{\text{other}} - |\text{rBias}|_{\text{NBD-MLE}})\,/\,|\text{rBias}|_{\text{other}}$.}
		\label{tab:rel_improve}
		\small
		\begin{tabular}{l c l c c c}
			\toprule
			Plot & $C$ (m) & Estimator & Q1 & Median & Q3 \\
			\midrule
			\multirow{8}{*}{BCI} 
			& 10 & $\hat{\lambda}_{C}^{(c)}$      & 0.231 & 0.659 & 0.858 \\
			& 10 & $\hat{\lambda}_{\text{DK}}$    & 0.820 & 0.882 & 0.940 \\
			& 20 & $\hat{\lambda}_{C}^{(c)}$      & 0.434 & 0.654 & 0.778 \\
			& 20 & $\hat{\lambda}_{\text{DK}}$    & -0.118& 0.515 & 0.889 \\
			& 30 & $\hat{\lambda}_{C}^{(c)}$      & 0.414 & 0.592 & 0.729 \\
			& 30 & $\hat{\lambda}_{\text{DK}}$    & -0.057& 0.338 & 0.654 \\
			& 40 & $\hat{\lambda}_{C}^{(c)}$      & 0.375 & 0.538 & 0.706 \\
			& 40 & $\hat{\lambda}_{\text{DK}}$    & -0.030& 0.336 & 0.526 \\
			\cmidrule{1-6}
			\multirow{8}{*}{HF}  
			& 10 & $\hat{\lambda}_{C}^{(c)}$      & 0.584 & 0.832 & 0.892 \\
			& 10 & $\hat{\lambda}_{\text{DK}}$    & -0.083& 0.656 & 0.884 \\
			& 20 & $\hat{\lambda}_{C}^{(c)}$      & 0.511 & 0.629 & 0.753 \\
			& 20 & $\hat{\lambda}_{\text{DK}}$    & -0.445& 0.395 & 0.591 \\
			& 30 & $\hat{\lambda}_{C}^{(c)}$      & 0.431 & 0.607 & 0.753 \\
			& 30 & $\hat{\lambda}_{\text{DK}}$    & 0.179 & 0.495 & 0.683 \\
			& 40 & $\hat{\lambda}_{C}^{(c)}$      & 0.397 & 0.636 & 0.756 \\
			& 40 & $\hat{\lambda}_{\text{DK}}$    & 0.305 & 0.512 & 0.743 \\
			\bottomrule
		\end{tabular}
	\end{table}
	
	
	\subsection{Censoring rates.}
\begin{figure}[!t]
		\centering
		\includegraphics[width=1\linewidth]{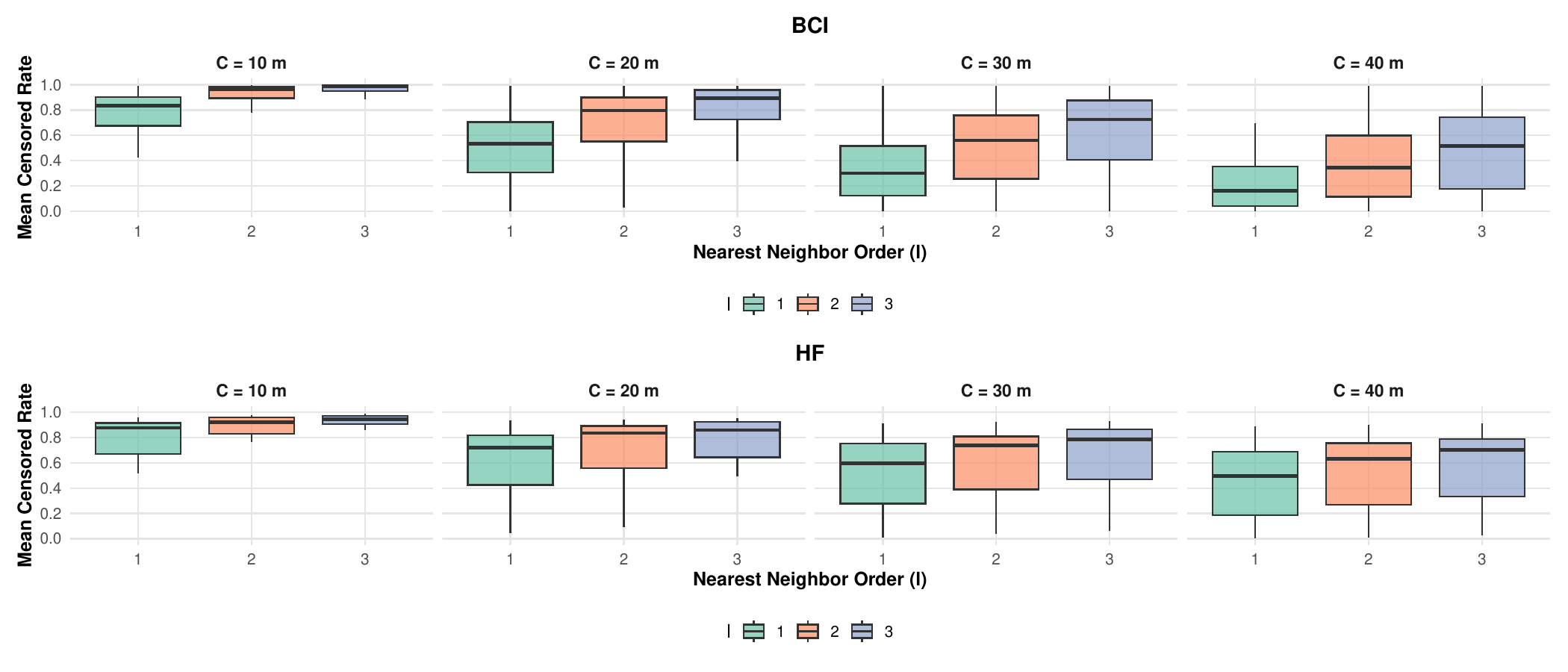}
		\caption{Distribution of mean censored rates under varying $\ell$ and $C$ for the BCI (upper) and Harvard Forest (lower) plots. Each boxplot summarizes the mean censored rate across 200 sampling replications for all analysed species (BCI: 112 species; Harvard Forest: 20 species; each with $\geq 500$ individuals).}
		\label{fig:mean_censored}
	\end{figure}

	Figure~\ref{fig:mean_censored} summarizes the mean censored rate (the proportion of sectors with fewer than $\ell$ individuals within $C$) for both forests. As expected, censoring increases markedly with higher $\ell$ and with smaller $C$, reinforcing the trade‑off discussed in the main text.
	
\end{document}